\documentclass[11pt]{article}
\usepackage{amsmath,amssymb}
\usepackage[dvips]{graphicx}
\usepackage{cite}
\begin{document}
\title{  Potential scatterings in the $L^2$ space :
 (2) Rigorous scattering probability of wave packets       }
\author{{Kenzo Ishikawa}$^{(1,2)}$,} 
\maketitle 
\begin{center}
(1)  Department of Physics, Faculty of Science, \\
Hokkaido University, Sapporo 060-0810, Japan \\
(2)Natural Science Center, \\
Keio University, Yokohama 223-8521, Japan\\
\end{center}


\begin{abstract}

In this study, potential scatterings are formulated in experimental setups with Gaussian wave packets in accordance  with a probability principle and associativity of products.   A breaking of an associativity is observed  in  scalar products with stationary  scattering states in  a majority of short-range potentials. Due to the breaking, states of different energies are not orthogonal and their superposition is not suitable for representing a normalized  isolate state. Free wave packets in perturbative expansions in coupling strengths keep the associativity, and give a rigorous 
amplitude that preserves manifest unitarity  and other principles of the quantum mechanics.    An absolute  probability  is finite and   comprises   cross sections and  new terms of unique properties. The  results also demonstrate  an interference term displaying  unique behavior  at an extreme forward direction.   
\end{abstract}
\newpage
\section{ Introduction }
Since the introduction of quantum mechanics,  particle scatterings  have been  expressed  with a cross section, which is partly  based on an analogy  of the scatterings of   electromagnetic waves and particles. In the quantum mechanics, waves  are connected with probabilities of physical processes and are not direct observables.   New observations \cite{EPR,aspect} have  suggested that the   transition process  in  quantum mechanics is not identical to that in  classical mechanics.   Therefore, study must explore  an accurate   scattering formula and  a precise  probability function. We  elucidate these topics   by using   a wave-packet formalism,  by which     a subtle problem of  nonorthogonality of scattering states of different energies  \cite{ishikawa_1} is resolved.  As such, we could determine  scattering probabilities comprising  various new terms in addition to  cross sections.

 	A  time-dependent Schr\"{o}dinger equation, i.e.,
\begin{eqnarray}
i \hbar \frac{\partial}{\partial t} \psi(t,\vec x)= (\frac{p^2}{2m} + V(\vec x))  \psi(t,\vec x), \label{schroedinger} 
\end{eqnarray}
satisfies a conservation of a probability density and current 
\begin{eqnarray}
& &\frac{\partial}{\partial t} \rho (t,\vec x)  + \nabla {\vec j(t,\vec x)}= 0, \label{conservation_l} \\
& &\frac{\partial}{\partial t} \int_{V} d{\vec x} \rho (t,{\vec x})  + \int_{\partial V}  d{\vec S} {\vec j(t,{\vec x})}= 0, \label{conservation_i}
\end{eqnarray}
where 
\begin{eqnarray}
& &\rho(t,\vec x)=  \psi(t,\vec x)^{*} \psi(t,\vec x), \nonumber \\
& &\vec j(t,\vec x)= \frac{i \hbar}{2m}(\psi(t, \vec x)^{*} ({\nabla} \psi(t,\vec x))-(\nabla \psi(t,\vec x))^{*} \psi(t,\vec x)).
\end{eqnarray}  
A cross section is a physical quantity   derived from rates, i.e., probabilities per  unit of time, \cite{goldberger, newton,taylor}. A standard method  utilizes   stationary solutions   $e^{\frac{E}{i\hbar }t}\psi_E(\vec x)$, where  $\psi_E(\vec x)$ denotes the eigenstates of a Hamiltonian 
\begin{eqnarray}
(\frac{p^2}{2m} + V(x)) \psi_E(\vec x)= E \psi_E(\vec x). 
\end{eqnarray}
The first term at the left-hand side of Eq.$(\ref{conservation_l})$ vanishes and the cross section is defined based on the  probability currents. 

 A transition amplitude  of a given   initial state to a certain final state is defined from the scalar product of these  states, and the square of the product represents its  probability. A particle is described by  a normalized state,  and can determine  probability in a unique manner.  Now, stationary scattering states are not normalized. As scalar products diverge, the probabilities are not uniquely computed. Normalized waves  are constructed easily from superpositions of stationary states. To represent physical states, norms of these states must be time-independent.  However,  stationary scattering states of different energies   are  known nonorthogonal  in most of potentials \cite{Landau-Lifshits}. Superposition of non-orthogonal states of different energies  have  time-dependent norms, and have  net probability currents in the spatial infinity as shown in  Eq.$(\ref{conservation_i})$.  These states do not express   isolated physical states even though they satisfy Schr\"{o}dinger's equation, and they are not suitable for  computing  rigorous  transition probabilities.

Therefore, to overcome the aforementioned  drawbacks of nonorthogonal scattering states, studies have proposed the adoption of formalisms using a quantization in a finite box,    regularized interaction $e^{-\epsilon |t|} V(x)$, or  rigged Hilbert space \cite{Landau-Lifshits,Schiff,Iwanami-kouza,Lipkin}. Although formalisms based on a finite box and the regularized interaction have been extensively applied in many areas,  these represent probabilities of idealistic systems, which are not identical to those of real systems.  
 In the third case, an extension of spaces is not straightforward and  has yet to be proven 
through  experiments.  In the current study, we focused on rigorous   scattering probabilities  in realistic  systems.

 To represent  an experimentally measured   probability,  formalism should be considered under both experimental and theoretical conditions. A transition probability of an initial state to   final state  is positive semidefinite and  smaller than  unity. These conditions are ensured  by normalized initial and final states. As Hermitian Hamiltonian  satisfies $H_{\alpha \beta}^{*}= H_{\beta \alpha}$ for normalized states $\alpha$ and $ \beta$, scattering amplitudes have been  shown to exist and  represent the amplitudes  in a Hilbert space \cite{Kato,Simon,kuroda}. However, deviations of this probability  from the conventional probabilities represented by only  the cross section have not yet been studied.

In this paper, we  present an absolute  transition probability for a potential scattering from wave-packet amplitudes  of normalized initial and final states.  We determined  various assertions on  a wave-packet formalism.

1. Associativity of products  is satisfied in numbers and matrices of finite dimensions, and it  is a fundamental relation that results in consistent  computations of physical quantities in  quantum mechanics. The associativity of operators and Hamiltonian  holds in a free system and in certain  potentials: however, it  does not  hold  in matrix elements of scattering states of short-range potentials.

2. In a functional space of stationary scattering states,  matrix manipulations could  deviate from those of  finite dimensions unless   a uniform convergence is achieved in an infinite dimension. Based on  matrix representations with a complete set of wave packets, we prove the existence of   non-uniform convergence  of operator products  of  stationary scattering  states in short-range potentials.  
Owing to this pathological property,  a  Hamiltonian becomes    not self-adjoint and  superpositions of  stationary states of different energies  display norms that  vary  in time. 
These wave functions demonstrate net probability currents at spatial infinity and  do not represent the scattering  of isolated particles, as shown in \cite{ishikawa_1}.

3. For short-range potentials, correct wave-packet amplitudes are displayed according to    time-dependent perturbative expansions in coupling strengths. A probability obtained in this manner satisfies  manifest unitarity and other principles of  quantum mechanics. This probability linearly depends on the  time interval and  is expressed by a slope, intercept, and  their  interference term.  An interference term  between the  zeroth-   and  first-order terms provides  new information  and may shed  a new light.  

Transition probabilities in terms of  probability principles in the  experimental setups have shown to be valuable   in many-body transitions \cite{ishikawa-shimomura-PTEP, ishikawa-tobita-PTEP,ishikawa-tobita-ann,ishikawa-tajima-tobita-PTEP,maeda-PTEP,ishikawa-oda-PTEP,ishikawa,ishikawa-nishiwaki-oda-PTEP}, and provide  fundamental  information 
  in potential scatterings. 

The remainder of this  paper is organized as  follows.  Second 2 summarizes the  wave packet representations, based on which we clarify  the  pathological behaviors of stationary scattering states.  In Section 3, we present the study of consistencies within wave packets with Hamiltonian dynamics in a coordinate representation.   Section 4 discusses  about the 
 wave-packet scattering phenomenon in  short-range potentials. In addition, the scattering amplitude and probability,  satisfying  the unitarity and principle of the quantum mechanics, are derived. Finally, we summarize the study in Section 5.  

\section{Wave-packet basis }

 Wave packets constructed from operators in a free theory are useful for  representing a Hamiltonian and other Hermitian operators.   In this representation,  a self-adjoint property of a space of scattering states is studied.      

Products of matrices of finite dimensions satisfy associativity. In matrices of infinite dimensions, the associativity is not necessary,  depending on the behaviors of 
convergences of matrix elements of the product, which are infinite series. This could be clarified through  two distinct cases. (1)  In the first case, the infinite  series are  uniformly convergent, and matrix products   preserve the  associativity.  The self-adjoint  operators  are represented by  self-adjoint operators in a space of infinite dimensions.    This is manifestly satisfied  in a space of bound states of the Hamiltonian  $\frac{p^2}{2m}+V(\vec x)$. 
(2) In another case, the series is not uniformly  convergent; thus, properties of finite dimensional matrices are not preserved, and  associativity does not hold and pathological behaviors are observed for  products of infinite dimensional matrices.   This occurs in stationary scattering states, and  a convergence becomes nonuniform. In addition, the  products of operators and the Hamiltonian are not associative in the majority of short-range potentials.

\subsection{ Matrix elements}
 A Gaussian wave packet in one spatial dimension of central values of momentum and positions $P_0$ and $X_0$  in a coordinate or momentum representation  is   defined by  matrix elements as follows:
\begin{eqnarray}
& &\langle x| P_0,X_0 \rangle=  N_1 e^{iP_0(x-X_0) -\frac{1}{2\sigma}(x-X_0)^2},    \label{coherent_packet}         \\
& &\langle p| P_0,X_0 \rangle=  N_1 \sigma^{1/2} e^{-ip X_0 -\frac{\sigma }{2}(p-P_0)^2},           \nonumber 
\end{eqnarray}
where $N_1^2= (\pi \sigma)^{-1/2}$. By substituting Eq.$(\ref{coherent_packet})$ into scalar products, we have 
\begin{eqnarray}
& &\langle P_1,X_1| P_2,X_2 \rangle = e^{-\frac{1}{4\sigma}(X_1-X_2)^2-\frac{\sigma}{4}(P_1-P_2)^2+i \frac{1}{2} (P_1+P_2)(X_1-X_2)},  \label{coherent_packet_2}\\
& &\langle P_0,X_0| P_0,X_0 \rangle = 1, \label{normalization}\\
& & \int_{-\infty}^{\infty} \frac{dP_0 dX_0}{(2 \pi)} | P_0,X_0 \rangle \langle  P_0,X_0| P_1,X_1 \rangle= |P_1,X_1 \rangle. \label{scalar_product_0}
\end{eqnarray}
The  relation Eq.$(\ref{scalar_product_0})$ is valid for arbitrary states $| P_1,X_1 \rangle $ and is equivalent to a completeness 
\begin{eqnarray}
 \int_{-\infty}^{\infty} \frac{dP_0 dX_0}{(2 \pi)} | P_0,X_0 \rangle \langle  P_0,X_0| = 1. \label{complete}
\end{eqnarray}
Accordingly, normalized states are expanded   as 
\begin{eqnarray}
& &| \Psi \rangle = \int_{-\infty}^{\infty} \frac{dP_0 dX_0}{(2 \pi)} | P_0,X_0 \rangle \langle  P_0,X_0| \Psi \rangle   \\
& &| \Phi \rangle = \int_{-\infty}^{\infty} \frac{dP_0 dX_0}{(2 \pi)} | P_0,X_0 \rangle \langle  P_0,X_0| \Phi \rangle,\end{eqnarray}
where matrix elements are given by
\begin{eqnarray}
\langle P_0,X_0 | \Psi \rangle= \int_{-\infty}^{\infty} dx \langle P_0,X_0| x \rangle \langle x|\Psi \rangle.
\end{eqnarray}
The  scalar product of two states is  expressed as
\begin{eqnarray}
\langle \Psi| \Phi \rangle&=& \int_{-\infty}^{\infty} \frac{dP_0 dX_0}{(2 \pi)}\int \frac{dP_0' dX_0'}{(2 \pi)} \langle \Phi |P_0 ,X_0 \rangle \langle P_0,X_0 | P_0',X_0' \rangle \langle  P_0',X_0'| \Psi \rangle\nonumber \\
 & =& \int_{-\infty}^{\infty} \frac{dP_0 dX_0}{(2 \pi)} \langle \Phi |P_0 ,X_0 \rangle \langle P_0,X_0 | \Psi \rangle,  
\end{eqnarray}
and 
\begin{eqnarray}
\langle \Psi| \Psi \rangle & =& \int_{-\infty}^{\infty} \frac{dP_0 dX_0}{(2 \pi)} |\langle \Psi |P_0 ,X_0 \rangle|^2  
\end{eqnarray}
where $\langle \Psi| \Psi \rangle $ is a positive semidefinite and vanishes only  when $ \langle \Psi|P_0, X_0 \rangle  = 0$ for arbitrary $P_0$ and  $X_0$. Consequently, if ${|\Psi \rangle= |\Phi \rangle}$,  ${\langle P_0,X_0|\Psi \rangle=\langle P_0,X_0 |\Phi \rangle}$ for arbitrary $P_0$ and  $X_0$. A set of  $|P_0, X_0 \rangle$ of arbitrary $P_0$ and  $X_0$ spans  nonorthogonal basis.

Matrix elements of the coordinate and its functions are denoted as,
\begin{eqnarray}
\langle P_1,X_1| x| P_2,X_2 \rangle& =&  N_1^2(\pi \sigma)^{1/2} a(1,2) \langle P_1,X_1| P_2,X_2 \rangle, \nonumber\\
\langle P_1,X_1| x^2| P_2,X_2 \rangle &=& ( \sigma/2+a(1,2)^2)   e^{ -\frac{1}{4\sigma} (X_1-X_2)^2+i\frac{P_1+P_2}{2}(X_1-X_2) -\frac{\sigma}{4}(P_1-P_2)^2},  \nonumber \\ 
 a(1,2)&=& \frac{( X_1 +X_2-i\sigma(p_1-p_2))}{2}, 
\end{eqnarray}
and
\begin{eqnarray}
& &\langle P_1,X_1| \delta(x) | P_2,X_2 \rangle = N_1^2    e^{i( P_1 X_1 -P_2X_2) -\frac{1}{2\sigma}( X_1^2+X_2^2)}.  \\
& &\langle P_1,X_1| e^{ikx} | P_2,X_2 \rangle = N_1^2  \sqrt{\sigma \pi}   e^{i \xi -\frac{1}{4\sigma}[ ( X_1-X_2)^2+\sigma^2(P_1-P_2-k)^2]}.\nonumber \\
& & \xi=\frac{1}{2}(P_1+P_2)(X_1-X_2)+\frac{k}{2}(X_1+X_2)
\end{eqnarray}

In addition, the matrix elements of the momentum are defined as
\begin{eqnarray}
\langle P_1,X_1| p| P_2,X_2 \rangle& =& N_1^2 \sigma b(1,2) (\frac{ \pi}{\sigma})^{1/2} 
 e^{ -\frac{1}{4\sigma} (X_1-X_2)^2+i\frac{P_1+P_2}{2}(X_1-X_2) -\frac{\sigma}{4}(P_1-P_2)^2}, \nonumber \\  
\langle P_1,X_1| p^2| P_2,X_2 \rangle &=& (\frac{1}{ 2 \sigma}+b(1,2)^2)  e^{ -\frac{1}{4\sigma} (X_1-X_2)^2+i\frac{P_1+P_2}{2}(X_1-X_2) -\frac{\sigma}{4}(P_1-P_2)^2},\nonumber  \\
b(1,2)&=&\frac{( P_1 +P_2+i\sigma^{-1}(X_1-X_2))}{2}. 
\end{eqnarray}
The matrices of operators are manifestly Hermitian.  Higher power terms are presented in the  Appendix.

\subsection{Products associativity  of operators : matrices of infinite dimensions }
In  finite-dimensional matrices, an element of the product of two matrices is uniquely calculated, and the matrix products satisfy an associativity.  However, in matrices  with an infinite dimension, the associativity depends on the behavior of an infinite dimensional  limit.   Let  $A$ and $B$ be matrices of infinite dimensions. Then 
an element of the matrix product    
\begin{eqnarray}
(AB)_{mn}=\lim_{L = \infty}\sum_{l= 1}^{l= L} A_{ml} B_{lm} \label{matrix_product}
\end{eqnarray}
is an   infinite series.  We derive two cases to determine the  behavior of  limit  $L \rightarrow   
 \infty$: 

(i)  A convergence of a series over $l$ up to $L$ is  uniform in $L \rightarrow \infty$, i.e., independent of $m$ and $n$. Then a large $L$ limit  can be  taken for arbitrary values of $m$ and $n$. 

(ii)  A convergence of a series  over $l$ up to $L$ is not uniform in $L \rightarrow \infty$, i.e., it depends on  $m$ and $n$. Then a large $L$ limit    of the matrix elements should be taken for each $m$ and $n$ separately. 

The products of the three matrices with  infinite dimensions.
\begin{eqnarray}
& &((AB)C)_{pq}=  \sum_{m= 1}^{\infty} \sum_{l= 1}^{\infty} A_{pl}B_{lm}C_{mq} \label{three_product_1}\\
& &(A(BC))_{pq}=\sum_{l= 1}^{\infty} \sum_{m= 1}^{\infty} A_{pl}B_{lm}C_{mq}\label{three_product_2}
\end{eqnarray}
always agree with each other in  case (i) but  not  in case (ii). In  case (ii), the product associativity is not ensured. 


A matrix element of an operator $O$ in two states $\psi(k)$ and $\psi(k')$ specified by continuous parameters $k$ and $k'$  is a set of integrals  of two  variable, namely $ \zeta_1 = (P_1,X_1)$ and $\zeta_2 = (P_2,X_2)$, respectively:   
\begin{eqnarray}
\langle \psi(k)| O | \psi(k')\rangle= \sum_{\zeta_i} \sum_{\zeta_j}  \langle \psi(k)| P_1,X_1 \rangle \langle P_1,X_1| O |  P_2,X_2 \rangle \langle P_2,X_2| \psi(k') \rangle, \label{matrix_element_u} 
\end{eqnarray}
where   $P$ and $X$ are continuous variables and  integrated up to   infinities:
\begin{eqnarray}
\sum_{\zeta}= \int_{-\infty}^{\infty} \frac{ dP dX}{2\pi}
\end{eqnarray}
Convergences  of the integrals  are equivalent to those of a series. An integration of  these variables with an order of $(i,j)= (1,2)$ conforms to that of  $(i,j)= (2,1)$ in  finite intervals, which in turn conforms to the case in which the integrals converge uniformly in an infinite interval. Conversely, if the integrals do not converge uniformly, this would not be true,and as such,  the associativity of  products would  not hold.  For an identity operator $O= 1$, Eq. $(\ref{matrix_element_u})$ becomes a scalar product 
\begin{eqnarray}
\langle \psi(k) | \psi(k')\rangle= \sum_{\zeta_i}   \langle \psi(k)| P_1,X_1 \rangle  \langle P_1,X_1| \psi(k') \rangle. \label{scalar_product_u} 
\end{eqnarray}
A behavior of the integrand at large $|\zeta|$ is  key for the  convergence of  integrals. Accordingly  the breaking of the associativity in Eq. $( \ref{matrix_element_u})$ and the 
nonorthogonality  in  Eq.$ ( \ref{scalar_product_u})$ have the  same origin.  
\subsubsection{Normalized states}
 For $\psi(k) = |P_i,X_i \rangle, \psi(k')=  |P_f,X_f \rangle $,
\begin{eqnarray}
& &large_{ |\zeta_1^M |} |\langle \psi(k)| P_1,X_1 \rangle \langle P_1, X_1| O |  P_2,X_2 \rangle| \leq O(e^{-({X_1}^M)^2-({P_1}^M)^2 }), \label{error_1}
\end{eqnarray}
where for a large $|\zeta_1| \gg |\zeta_2|$, the right-hand side of Eq. $(\ref{error_1}) $ is arbitrarily small and independent of $X_2$ and $P_2$. Accordingly, 
\begin{eqnarray}
 & & lim_{|\zeta_1^M| \rightarrow \infty}  \sum_{\zeta_1= -\zeta_1^M}^{\zeta_1= \zeta_1^M}\langle \psi(k)| P_1,X_1 \rangle \langle P_1, X_1| O |  P_2,X_2 \rangle =  \langle \psi(k)|  O|  P_2,X_2 \rangle ,\nonumber  \\
& & 
\end{eqnarray}
is uniformly convergent.

Similarly,
\begin{eqnarray}
 &large_{|\zeta_2^M|}\langle P_1,X_1| O |  P_2,X_2 \rangle \langle P_2,X_2| \psi(k') \rangle| \leq 
 O(e^{-( ( X_1^M)^2 +(X_2^M)^2)}), \label{error_2}
\end{eqnarray}
where for a large $|\zeta_1| \gg |\zeta_2|$, the right-hand side of Eq.$(\ref{error_2}) $ is arbitrarily small and independent of $X_2$ and $P_2$. Accordingly, 
\begin{eqnarray}
& & lim_{|\zeta_2^M |\rightarrow \infty}\sum_{X_2=-X_2^M}^{X_2=X_2^M} \langle P_1,X_1| O |  P_2,X_2 \rangle \langle P_2,X_2| \psi(k') \rangle = \langle P_1,X_1| O | \psi(k') \rangle \nonumber \\
& & 
\end{eqnarray}
is uniformly convergent. For bound  states, $\psi(k)$ is a normalized superposition of  $|P_i,X_i \rangle$, and the convergences  of the integrals  are uniform. 
Then, matrix elements satisfy 
\begin{eqnarray}
& &\sum_{\zeta_1}   \langle \psi(k)| P_1,X_1 \rangle \sum_{\zeta_2} \langle P_1,X_1| O|  P_2,X_2 \rangle \langle P_2,X_2| \psi(k') \rangle \nonumber \\
& &=\sum_{\zeta_2}   \langle \psi(k)| P_1,X_1 \rangle \sum_{\zeta_1}\langle P_1,X_1| O |  P_2,X_2 \rangle \langle P_2,X_2| \psi(k') \rangle.
\end{eqnarray}
\subsubsection{Non-normalized states }
For stationary scattering states, $\psi(k)$ is a non-normalized superposition of  $|P_i,X_i \rangle$, and it includes an infinitely large value of $\zeta_i$. The convergences  of the integrals may differ  from those of   normalized states, and they satisfy
\begin{eqnarray}
& &large_{ |\zeta_1^M |} |\langle \psi(k)| P_1,X_1 \rangle \langle P_1, X_1| O |  P_2,X_2 \rangle| \leq O(e^{-({X_1}^M)^2-({P_1}^M)^2 }). \label{error_{n1}}
\end{eqnarray}
Now for a large value of $|\zeta_2|$, the right-hand side of Eq. $(\ref{error_{n1}}) $ is finite at   $X_1 \approx X_2$ and $P_1 \approx P_2$; this  cannot be  chosen to be independent of $X_2$ and $P_2$. A limit 
\begin{eqnarray}
 & & lim_{|\zeta_1^M| \rightarrow \infty}  \sum_{\zeta_1=-\zeta_1^M}^{\zeta_1=\zeta_1^M}\langle \psi(k)| P_1,X_1 \rangle \langle P_1, X_1| O |  P_2,X_2 \rangle =  \langle \psi(k)| O |  P_2,X_2 \rangle ,\nonumber  \\
& & 
\end{eqnarray}
may not be   uniformly convergent.
Similarly,
\begin{eqnarray}
& &large_{|\zeta_2^M|}\langle P_1,X_1| O |  P_2,X_2 \rangle \langle P_2,X_2| \psi(k') \rangle| \leq 
 O(e^{-( ( X_1^M)^2 +(X_2^M)^2)}), \label{error_{n2}} \nonumber\\
& &
\end{eqnarray}
where  the right-hand side of Eq. $(\ref{error_{n2}})$ depends on $X_1$ and $P_1$. If this cannot be  chosen to be independent of $X_1$ and $P_1$, the following limit is not uniformly convergent:
\begin{eqnarray}
& & lim_{|\zeta_2^M |\rightarrow \infty}\sum_{X_2= -X_2^M}^{X_2= X_2^M} \langle P_1,X_1| O |  P_2,X_2 \rangle \langle P_2,X_2| \psi(k') \rangle = \langle P_1,X_1| O || \psi(k') \rangle. \nonumber \\
& & 
\end{eqnarray}
In addition, the convergences of the integrals are not uniform. For stationary scattering   states, $\psi(k)$ and $\psi(k')$ are   non-normalized superpositions of  $|P_i,X_i \rangle$. Here, we show that the convergences  of the integrals are not uniform, and the  matrix elements satisfy 
\begin{eqnarray}
& &\sum_{\zeta_1}   \langle \psi(k)| P_1,X_1 \rangle \sum_{\zeta_2} \langle P_1,X_1| O |  P_2,X_2 \rangle \langle P_2,X_2| \psi(k') \rangle \nonumber \\
& & \neq \sum_{\zeta_2}   \langle \psi(k)| P_1,X_1 \rangle \sum_{\zeta_1}\langle P_1,X_1| O |  P_2,X_2 \rangle \langle P_2,X_2| \psi(k') \rangle.  
\end{eqnarray}

\subsubsection{Scalar products of stationary states }
We examine the associativity according to explicit calculations. An eigenstate in the x-space of a particle of mass $m$  for   a case of parameters $m= 1$  is presented. Then, the solutions can be expressed as
  \begin{eqnarray}
\psi_k(x)&= &e^{ikx}+ R(k) e^{-ikx} ;x< 0 \label{delta_potential_1}\\
&=&  T(k) e^{ikx} ;0< x\nonumber 
\end{eqnarray}
where for the potential $V(x)=  g \delta (x)$,
\begin{eqnarray}
R(k)= \frac{-ig}{k+ig}, T(k)= \frac{k}{k+ig}\label{delta_potential_2}.
\end{eqnarray} 
 are studied.

In the wave packet representation,  the state in Eq. $( \ref{delta_potential_1})$ is expressed as 
\begin{eqnarray}
& &\langle P,X | \psi_k \rangle = \langle P,X |x \rangle \langle x|  \psi_k \rangle \nonumber \\
& &= N_1[\int_{-\infty}^0 dx ( e^{ikx}  \tilde I_0^{*} (P,X) + R(k) e^{-ikx} \tilde I_0^{*}(P,X))+ 
  \int_0^{\infty} dx e^{ikx} T(k) \tilde I_0^{*}(P,X) ], \nonumber \\ \label{stationary_state}
\end{eqnarray}
where
\begin{eqnarray}
& &\tilde I_0^{*}(P,X)=  e^{-iP(x-X) -\frac{1}{2\sigma}(x-X)^2}.
\end{eqnarray}
By using error functions, Eq. $(\ref{stationary_state})$   is written as
\begin{eqnarray}
& &N_1(\pi \sigma)^{1/2}[ e^{ikX-\frac{\sigma}{2}(k-P)^2} \pi \sqrt{\sigma}+ R(k) e^{-ikX-\frac{\sigma}{2}(k+ P)^2} \sqrt{\frac{\sigma}{\pi}}( 1-erf(\frac{X+i\sigma(k+ P)}{\sqrt{ 2 \sigma}}))\nonumber \\
& &+(T(k)-1)e^{ikX-\frac{\sigma}{2}(k-P)^2} \sqrt{\frac{\sigma}{\pi}}( 1-erf(\frac{-X+i\sigma(k-P)}{\sqrt{ 2 \sigma}}))]. \label{wave_stationary}
\end{eqnarray}
By substituting an  asymptotic behavior of the error function
\begin{eqnarray}
erf(z)=sgn(z)-\frac{1}{{\sqrt \pi} z}e^{-z^2}+ \cdots,
\end{eqnarray}
we determine the behavior of  the second term  in the right-hand side as 
\begin{eqnarray}
 & &e^{-ikX -\frac{\sigma}{2}((k+ P)^2}(\sqrt{\frac{\sigma \pi}{2}})^2 (1-erf(\frac{X+i\sigma(k+ P )}{\sqrt{2 \sigma}})) \nonumber \\ 
& &= e^{-ikX -\frac{\sigma}{2}((k+ P)^2}(\sqrt{\frac{\sigma \pi}{2}})^2 (1-sgn(X)) \nonumber \\
& &+
       e^{-ikX }((\sqrt{\frac{\sigma \pi}{2}})^2) \frac{\sqrt{2 \sigma}}{{\sqrt \pi} (X+i\sigma(k+P))}e^{-\frac{(X^2+2i X\sigma(k+ P ))}{2 \sigma}} , \label{wave_function_w}
\end{eqnarray}
where the first term decreases exponentially with respect to  momentum $P$  and the second term decreases marginally with respect to $P$.  The third term  in the right-hand side of
 Eq. $(\ref{stationary_state})$ is written in a similar manner. 

By substituting Eq.$(\ref{wave_stationary})$ in  Eq. $(\ref{scalar_product_u})$, we have the scalar product of two states  as 
\begin{eqnarray}
& &\langle \psi_k| P,X \rangle \langle P,X | \psi_{k'} \rangle= N_1^2(\pi \sigma) \int \frac{dP dX}{2\pi} \nonumber \\
& &~~~~~~  [\tilde A_{00}+  R(K)^{*}  R(k') \tilde A_{11}+ (T(k)^{*}-1)(T(k')-1) \tilde A_{22}+\cdots]  \nonumber \\
& & 
\end{eqnarray}
where  
\begin{eqnarray}
& &\tilde A_{00} =  e^{-i(k-k')X -\frac{\sigma}{2}((k-P)^2+ (k'+ P)^2)}(\sqrt{\frac{\sigma \pi}{2}}2 \sqrt \pi)^2\\
& &\tilde A_{11}=  e^{i(k-k')X -\frac{\sigma}{2}((k-P)^2+(k'+P)^2)}(\sqrt{\frac{\sigma \pi}{2}})^2 (1-erf^{*}(\frac{X+i\sigma(k-P )}{\sqrt{2 \sigma}}))\nonumber \\
& &~~~~~~\times ( 1-erf(\frac{X+i\sigma(k'+P )}{\sqrt{2 \sigma}} )) \\
& &\tilde A_{22}=  e^{-i(k-k')X -\frac{\sigma}{2}((k-P)^2+(k'+P)^2)}(\sqrt{\frac{\sigma \pi}{2}})^2 (1-erf^{*}(\frac{-X+i\sigma(k-P )}{\sqrt{2 \sigma}}))\nonumber \\
& &~~~~~~\times ( 1-erf(\frac{-X+i\sigma(k'+P )}{\sqrt{2 \sigma}} ))  
\end{eqnarray}
The integrals 
and other terms are denoted as   $\cdots$, and have  not been provided  here for simplicity. 

An integration over the phase space achieves  a marginal contribution from the  second term in the right-hand of the equation.  The details are  given in Appendix D. It follows 
\begin{eqnarray}
& &\tilde A_{22}(P,X)=  \tilde A_{11}(-P,-X), 
\end{eqnarray}
and  integrations over the phase space 
\begin{eqnarray}
& &\int \frac{dP dX}{2\pi} \tilde A_{00} =  \delta(k-k') \int dP e^{-{\sigma}(k-P)^2}(\sqrt{\frac{\sigma \pi}{2}}2 \sqrt \pi)^2,\\
& &\int \frac{dP dX}{2\pi}\tilde A_{11}= \int \frac{dP dX}{2\pi}\tilde A_{22} \nonumber \\
& &= 2\sigma^2  \int \frac{dP dX}{2\pi} e^{i(k-k')X -\frac{( X^2+i X\sigma(k'-k ))}{ \sigma}} \frac{1}{ (X-i\sigma(k+P))}   \frac{1}{ (X+i\sigma(k'+P))}. \nonumber     \\
\end{eqnarray}
The integration of $\tilde A_{00}$ is proportional to $\delta(k-k')$, and those of $\tilde A_{11}$ and $\tilde A_{22}$ are not proportional to $\delta(k-k')$ but have contributions from  marginal terms. 
The marginal terms in Eq. $(\ref{wave_function_w})$ result in a  nonvanishing nondiagonal term  $k \neq k'$.

\subsection{Matrix element of operator }
In the stationary scattering state, product  associativity   in matrix elements of an operator is broken.  By substituting Eq. $(\ref{wave_stationary})$ in  Eq. $(\ref{matrix_element_u})$, the matrix element can be written as   
\begin{eqnarray}
& &\sum_{\zeta_i} \sum_{\zeta_j}  \langle \psi(k)| P_1,X_1 \rangle \langle P_1,X_1| O |  P_2,X_2 \rangle \langle P_2,X_2| \psi(k') \rangle  \nonumber \\
& &\langle \psi_k| P_1,X_1 \rangle \langle P_1,X_1 |O|P_2,X_2 \rangle \langle P_2,X_2| \psi_{k'} \rangle=N_1^4(\pi \sigma)^2 \int \frac{dP_1 dX_1 dP_2 dX_2}{(2\pi)^2} \nonumber \\
& &~~~~~~  [\tilde B_{00}+R(k)^{*}  R(k') \tilde B_{11}+(T(k)^{*}-1)(T(k')-1) \tilde B_{22}+\cdots] \label{operator_m} \nonumber \\
& & 
\end{eqnarray}

For an opetator $O=1+p$,  $\tilde B_{00}$ and the  marginal term of $\tilde B_{11}$ can be formulated as     
\begin{eqnarray}
\tilde B_{00} &=& e^{-i(k-k')X -\frac{\sigma}{2}((k-P_1)^2+(k'+P_2)^2)}(\sqrt{\frac{\sigma \pi}{2}}2 \sqrt \pi)^2\langle P_1,X_1|O |P_2,X_2 \rangle, \\
\tilde B_{11}&=& e^{ikX_1} \sqrt{\frac{\sigma}{\pi}} \frac{\sqrt{2 \sigma}}{{\sqrt \pi} (X_1-i\sigma(k+P_1))}e^{-\frac{(X_1^2-2i X_1\sigma(k+P_1 ))}{2 \sigma}} \nonumber \\
& &\langle P_1,X_1|O |P_2,X_2 \rangle   e^{-ik'X_2} \sqrt{\frac{\sigma}{\pi}} \frac{\sqrt{2 \sigma}}{{\sqrt \pi} (X_2+i\sigma(k'+P_2))}e^{-\frac{(X_2^2+2i X_2\sigma(k'+P_2 ))}{2 \sigma}}                   
,
\nonumber \\
& &=N_{11}\frac{e^{-f_{11} }}{ (X_1-i\sigma(k+P_1))}\frac{1}{ (X_2+i\sigma(k'+P_2))} [1+\frac{1}{2}(P_1+P_2+i \frac{X_1-X_2}{\sigma})],\nonumber \\
\end{eqnarray}
where $f_{00}$  is a sum of  $i(k-k')(X_1+ X_2)/2$ and a function of $X_1-
X_2$ and the integral over the positions of $\tilde B_{00}$ is proportional to $\delta(k-k')$. 

Then,  exponential factor $f_{11}$ is given as
\begin{eqnarray}
& &f_{11}=-i(kX_1-k'X_2)+\frac{(X_1^2-2i X_1\sigma(k+P_1 ))}{2 \sigma}+\frac{(X_2^2+2i X_2\sigma(k'+P_2 ))}{2 \sigma}\nonumber \\
& &+\frac{1}{4\sigma} (X_1-X_2)^2 +\frac{\sigma}{4}(P_1-P_2)^2 -i \frac{P_1+P_2}{2}(X_1-X_2),
\end{eqnarray}
and the matrix element is written as 
\begin{eqnarray}
\langle P_1,X_1|O |P_2,X_2 \rangle& =&N_1^2 (\pi \sigma)^{1/2} e^{-\frac{1}{4\sigma} (X_1-X_2)^2 -\frac{\sigma}{4}(P_1-P_2)^2 +i \frac{P_1+P_2}{2}(X_1-X_2)} \nonumber \\ 
& \times &[1+\frac{1}{2}(P_1+P_2+i \frac{X_1-X_2}{\sigma})]. 
\end{eqnarray}

To  integrate variables  $X_1$ and $P_1$,  the integrand is written in their  diagonal form,  
considering   central values  $X_1^0$ and $P_1^0$ of the  functions of $X_2$ and $P_2$, respectively. The Integrations over $X_2$ and $P_2$ are  easily achieved  by writing   exponents in the following respective   diagonal forms,
\begin{eqnarray}
& &f_{11} =  \frac{1}{2\sigma}[ { \frac{3}{2}(X_1-X_1^0)^2+2\sigma (P_1-P_1^0)^2}+\frac{4}{3} (X_2-X_2^0)^2+2 \sigma(P_2-P_2^0)^2 ] \nonumber \\
& &X_1^0= \frac{3}{8}X_2+i \frac{3}{8}\sigma(3P_1+P_2), P_1^0=0, X_2^0=-i\sigma P_2    ,P_2^0=0
\end{eqnarray}
where $X_1^0$ and $P_1^0$ are functions of $X_2$ and $P_2$, respectively. Note that  the coefficients were obtained as  $\frac{3}{2}$  and $\frac{4}{3}$ for $(X_1-X_1^0)^2$ and $(X_2-X_2^0)^2$ respectively. These coefficients depend on the order of integrations.  Now a denominator of the  marginal terms depends on $k$ and another depends on $k'$, and  the coefficients of $(X_1-X_1^0)^2$ differ  from  the coefficients of $(X_2-X_2^0)^2$ in $f_{11}$. Accordingly, the contributions form an integration over the phase space depending  on the order of the intergratons. This assertion is valid for a general  operator $O$. 
It is sufficient to establish   that the associativity is broken because of the nonassociative nature  one term $B_{00}$ of the matrix element for $k \neq k'$.    

The formulae for the potential  $\delta(x)$ presented here are valid  to the margnal terms  in  genaral short-range potentials. See   Appendix E. 

\subsection{Eigenstates of  Hamiltonian }
This section presents a  system in which   the associativity is not  satisfied.  Here, the   Hamiltonian   is expressed as  
\begin{eqnarray}
& &\langle P_1,X_1|H| P,X \rangle = \frac{1}{2m} \langle P_1,X_1|p^2| P,X \rangle + g \langle P_1,X_1|V(x)| P,X \rangle \nonumber  \\
& &= \frac{1}{2m} (\frac{1}{ 2 \sigma}+b(1,2)^2)  e^{ -\frac{1}{4\sigma} (X_1-X)^2+i\frac{P_1+P}{2}(X_1-X) -\frac{\sigma}{4}(P_1-P)^2}+g N_1^2   V(P_1,P;X_1,X), \nonumber \\
& & \label{hamiltonian}
\end{eqnarray}
where for $V(x)=  \delta(x) $,
\begin{eqnarray}
& &V(P_1,P,X_1,X)=  e^{i( P_1 X_1 -PX) -\frac{1}{2\sigma}( X_1^2+X^2)}, \label{short_range}
\end{eqnarray}
and for $V(x)=  (\frac{4\pi}{\sigma_V})^{1/2}e^{-\frac{1}{\sigma_V}(x-X_V)^2 }$ ,
\begin{eqnarray}
 & &V(P_1,P,X_1,X)=  V_0  e^{-\frac{1}{4\sigma}(X_1-X)^2-\frac{\sigma}{4}(P_1-P)^2 +\zeta } \nonumber \\
& &\zeta=\frac{\sigma^2}{4(\sigma_V+\sigma)}(P_1-P)^2-\frac{1}{\sigma_V+\sigma} (\frac{X_1+X}{2}-X_V)^2 +\frac{i}{2}(P_1+P)(X_1-X) \nonumber \\
& &-i\frac{\sigma}{\sigma_V+\sigma}(P_1-P)(\frac{X_1-X}{2}-X_V) 
. \label{G_potential} 
\end{eqnarray}
where $V_0$ is a constant which depends on wave packets and the potential.   
The first term in the right hand side of Eq.$(\ref{hamiltonian})$ decreases rapidly with $P_1-P$ and $X_1-X$. The second term  for the Gaussian potential is slightly modified but also  decreases rapidly with  $P_1-P$ and $X_1-X$. The second term   for the  potential $\delta(x)$  decreases rapidly with  $X_1$ and $X$. 
 \subsubsection{Eigenvalue equation}
In the wave-packet representation, an eigenvalue equation is formulated as
\begin{eqnarray}
& &\sum_{P,X} \langle P_1,X_1|H| P,X \rangle\langle P,X | \psi_k \rangle = E_k \langle P_1,X_1|\psi_k \rangle.  \label{eigenvalue_eq}
\end{eqnarray}

Note that  the integral over phase space $P,X$ in Eq. $(\ref{eigenvalue_eq})$ must converge uniformly, independent of the final values $P_1,X_1$. If this is not true,  several trivial properties of finite-dimensional matrices, such as  an orthogonality of two states of different eigenvalues are  not satisfied. 

{ \bf Theorem} :

 For $ X_1 \rightarrow \infty $, $ \langle P_1,X_1|\psi_k \rangle \neq 0$ depending on $X_1$. Then the convergence of a summation over $P,X$ on the left-hand side  of Eq. $(\ref{eigenvalue_eq})$ depends on $X_1$ and is not uniform.  

Proof. For a  potential presented in Eq. $(\ref{G_potential}) $, $ \langle P_1,X_1|H| P,X \rangle $ is finite only in the region of the finite $|X_1-X|$ for large $X_1$ and $X$.  The  region in $X$ is not fixed but depends on  $X_1$. Hence the convergence of the series in $X$  is not uniform. The products do not satisfy the associativity generally.   
For the  potential $\delta(x)$ shown in Eq. $(\ref{short_range}) $, $ \langle P_1,X_1|V| P,X \rangle $ demonstrates  a slightly modified property, and it is finite in the region of finite $|X_1-X|$ for large values of $X_1$ and $X$  in the region of finite $|X|$. However,  the Hamiltonian  $ \langle P_1,X_1|H| P,X \rangle $ is finite only in the region of the finite $|X_1-X|$ for large $X_1$ and $X$.  Hence the convergence of the series in $X$  is  non-uniform.  

This theorem is extended to matrix elements of another operator  $O(x,p)$,
 \begin{eqnarray}
& &\sum_{P,X} \langle P_1,X_1|O| P,X \rangle\langle P,X | \psi_k \rangle  \label{Operator}
\end{eqnarray}
where $ \langle P_1,X_1|O| P,X \rangle $ is finite only in the region of finite $|X_1-X|$ for large $X_1$ and $X$ values.
The  region in $X$ is not fixed but depends on  $X_1$. Hence the convergence of the series in $X$  is not uniform, and the products do not satisfy the associativity generally.

\subsubsection{Test of associativity: on order of the products  }
The complex conjugate of the eigenvalue equation  Eq. $(\ref{eigenvalue_eq})$ is formulated as 
\begin{eqnarray}
& &(\sum_{P,X} \langle P_1,X_1|H| P,X \rangle\langle P,X | \psi_k \rangle )^{*}=  E_k^{*} (\langle P_1,X_1 | \psi_k \rangle)^{*}, \\
& &\sum_{P,X} \langle \psi_k|P,X \rangle \langle P,X|H^{\dagger}|P_1,X_1 \rangle = E_k^{*} \langle \psi_k|P_1,X_1 \rangle.  
\end{eqnarray}
Here,  notations $k \rightarrow k',(P,X) \rightarrow (P_1,X_1)$, can be changed; then,
\begin{eqnarray}
& &\sum_{P_1,X_1} \langle \psi_{k'}|P_1,X_1 \rangle \langle P_1,X_1|H^{\dagger}|P,X \rangle = E_{k'}^{*} \langle \psi_{k'}|P,X \rangle.  \label{eigenvalue_eq_2}
\end{eqnarray}

In Eq. $(\ref{eigenvalue_eq})$,  $ \langle \psi_{k'} |  P_1,X_1 \rangle $ is multiplied to the left hand side and  a summation is taken  over ${P_1,X_1}$ to get
\begin{eqnarray}
& & \sum_{P_1,X_1}\langle \psi_{k'} |  P_1,X_1 \rangle  \sum_{P,X} \langle P_1,X_1|H| P,X \rangle\langle P,X | \psi_k \rangle =  E_k  \sum_{P_1,X_1} \langle \psi_{k'} |P_1,X_1 \rangle  \langle P_1,X_1 | \psi_k \rangle. \nonumber \\
& &\label{eigenvalue_eq_3}
\end{eqnarray}

Next   $\langle P,X|\psi_k \rangle$ is multiplied to the  right of Eq. $(\ref{eigenvalue_eq_2})$, and  a summation is taken  over ${P,X}$
\begin{eqnarray}
& & \sum_{P,X}  \sum_{P_1,X_1} \langle \psi_{k'} |  P_1,X_1 \rangle  \langle P_1,X_1|H| P,X \rangle\langle P,X | \psi_{k} \rangle =  E_{k'}^{*}  \sum_{P,X} \langle \psi_{k'} |P,X \rangle  \langle P,X | \psi_{k} \rangle \nonumber \\
& & \label{eigenvalue_eq_4}
\end{eqnarray}

The order of the summations in the  Eqs. $(\ref{eigenvalue_eq_3})$ and $(\ref{eigenvalue_eq_4})$ are reversed. As the limits to infinite dimensions are not uniformly convergent, these cannot be  guaranteed to be equivalent.
This satisfies 
\begin{eqnarray}
(E_k -E_k^*) \sum_{P_1,X_1} \langle \psi_{k'} |P_1,X_1 \rangle  \langle P_1,X_1 | \psi_k \rangle \neq 0,
\end{eqnarray} 
and the two states are not orthogonal for $E_k \neq E_{k'}$.  

Non-associativity of the matrix products of the scattering states results in   
eigenstates of different eigenvalues to not  be orthogonal in general.  Here, the  orthogonality satisfied in finite-dimensional matrices was   violated in the  infinite-dimensional matrices.  This pathological property is a general feature of stationary states.  

\subsubsection{Summary on self-adjoint properties }
As the wave packets are Gaussian in $x$ and $p$, and they  vanish rapidly at $|x|, |p| \rightarrow \infty $, the operators defined from these variables are manifestly Hermitian in the wave-packet representations. Accordingly, the free Hamiltonian, $H_0$, and the total Hamiltonian, $H$, are manifestly Hermitian in the wave-packet representation. The product associativity is  satisfied in the normalized states,  but  broken in the scattering states,  with respect to the  eigenstates  of the Hamiltonian of the continuum spectrum.  The matrix element of operators in the scattering states violates the associativity of products. As standard calculations are made under associativity, the associativity is considered inevitable.

The next section presents a formulation of the scattering amplitude and probability  without the consideration of  non-associativity of products.

\section{Consistency of products in coordinate representation  }
Here, we prove then  violation of the associativity based on a contradiction. In a system where  the associativity of  products is not satisfied for some matrix elements of operators $A$,$B$, and $C$
\begin{eqnarray}
(A(BC)-B(AC))_{lm} \neq( (AB)C-(BA)C)_{lm}=([A,B]C)_{lm}, \label{associativity_commutation}
\end{eqnarray}
its calculation should result  a contradiction if the right-hand side is substituted for the left-hand side.   We confirm that the matrix elements  derived from ignoring  the breaking of associativity are in contradiction.  
    
In this section, we use the following  regularized integral:
\begin{eqnarray}
\int_0^{\infty} dx e^{i(k+ i\epsilon)}=\frac{i}{(k+ i \epsilon)}. 
\end{eqnarray}

\subsection{Orthogonality of stationary states }
The eigenstates of different eigenvalues of a Hermitian Hamiltonian $H= H_0+ V$,( where $H_0= 
 \frac{p^2}{2m}$ and $V$ is a potential), are orthogonal in a Hilbert space. In a short-range potential, states are expressed by the  continuum parameter $k$. Let $\psi_k(x)$ and $\psi_{k'}(x)$ be eigenstates of  $H$  satisfying $H^{\dagger} = H$ with energies $E(k)$ and $E(k')$, respectively. Then, the corresponding scalar products $ \langle \psi_k' | \psi_k \rangle$ and $ \langle \psi_k' | H|\psi_k \rangle$ can be given as
\begin{eqnarray}
& &I_0(k',k)= \int_{-\infty}^{\infty} dx( \psi_{k'}^{*} (x)  \psi_{k}(x) ) , E(k',k)= \int_{-\infty}^{\infty} dx( \psi_{k'}^{*} (x)|H|  \psi_{k}(x) ). \nonumber \\
& &
\end{eqnarray}
These are assumed to be finite. Herein the upper and lower bounds of an integral are  $ -\infty$ and $\infty$, respectively, which   are omitted.   When $H$ is first multiplied to state $\psi_k(x)$, we have  
\begin{eqnarray}
& &E(k',k)= E(k) I_0(k',k), \label{m_energy_1} 
\end{eqnarray}
and when $H$ is first multiplied to  state $\psi_{k'}(x)$, then 
\begin{eqnarray}
& &E(k',k)=  E(k') I_0(k',k). \label{m_energy_2}
\end{eqnarray}

 For $A= \psi(k')$,  $B= H$,  and   $C= \psi(k)$  in  (i) of Eqs.$( \ref{three_product_1})$ and $(\ref{three_product_2})$, i.e., the limit is uniformly convergent,  Eq.$(\ref{m_energy_1})$ conforms to  Eq. $(\ref{m_energy_2}) $. This equality  can be  written as 
\begin{eqnarray}
& &(E(k')-E(k)) I_0(k',k) = 0.
\end{eqnarray}
Accordingly, the two states satisfy
\begin{eqnarray}
 I_0(k',k)= \langle \psi(k')| \psi(k) \rangle= 0 ~\text{for} ~ E(k) \neq E(k'). \label{orthogonality}
\end{eqnarray}
As observed, the states with  different energies are orthogonal.  

  In  (ii) of Eqs.$( \ref{three_product_1})$ and $(\ref{three_product_2})$, i.e., a convergence of a summation over $l$ is not uniform, Eq. $(\ref{m_energy_1}) $ does not conform  to Eq.$(\ref{m_energy_2}) $, and
\begin{eqnarray}
& &(E(k')-E(k)) I_0(k',k)  \neq 0,
\end{eqnarray}
Therefore, the states with  different energies are not orthogonal.  

From Sub-section 2.4, stationary states  in a short-range potential of finite width  belong to the category (ii), and  are not orthogonal. For a potential $\delta(x)$, the matrix element behaves marginally, and is studied   further in detail in the following.

\subsection{Consistency of wave packets with Hamiltonian dynamics: commutation relations}
The products of three matrices are associative and   Eq. $(\ref{three_product_1})$ conforms  to  Eq. $(\ref{three_product_2})$ in the case (i) of Eqs.$( \ref{three_product_1})$ and $(\ref{three_product_2})$ where the  convergence of a summation over $l $ or that over $m$ is uniform; however,  these do not conform  in (ii)  of Eqs.$( \ref{three_product_1})$ and $(\ref{three_product_2})$ where the  convergence of a summation over $l$ or $m$ is not uniform. 
Matrix elements of product of  $A$ that  characterizes Gaussian wave packets and a Hamiltonian are studied  for various  potentials.  Here,   unit $\hbar=1$ is used.

Let us define a wave packet satisfying minimum uncertainty, coherent state, with  operator $A$  as
\begin{eqnarray}
& &A^{\dagger} |\psi_c \rangle = 0, A= p+ ix, \label{coherent_state}\\
& &[x,p]= i (\hbar=1).
\end{eqnarray}
Then, the coherent state is  stable when  this relation is satisfied independent of time.   

\subsubsection{ Evolution operators} 

 In a system described by the Hamiltonian of $m= 1$, 
\begin{eqnarray}
& & H= H_0+ V(x), H_0= p^2/2,  
 \end{eqnarray}
we define an operator $A_n$ from a commutation relation 
\begin{eqnarray}
A_n=[[[[A,H ],\cdots ,H], H] \label{commutationH}
\end{eqnarray}
which can be written as   
\begin{eqnarray}
A_0= A, A_{n}-[A_{n-1},H]= 0,  n \geq 1. \label{recursion_1}
\end{eqnarray}

By assuming product associativity for these  operators,  we have the commutation relations $[A_n,H]$ as   
\begin{eqnarray}
A_1&=&  -p -i  V'(x),\label{A_1} \\
A_2&=&   \frac{1}{2}\left( V''(x)   p+   p V''(x)  +  i V'(x) \right). \label{A_2}  \\
A_{3 }&=& [A_2,H] = i\frac{1}{2^2} \left(p^2 V'''(x) + 2p V'''(x)+ V'''(x) p^2 \right) \nonumber  \\
   & &+ i\frac{1}{2^2}\left(+ ip V''(x) - iV''(x) p  \right)- i V'(x)V''(x),  \label{A_3}
\end{eqnarray}
where  $A_1$ and $A_2$ are linear  derivatives of the potential, and $A_3$ is bi-linear. Furthermore,   $A_n,(n>3)$ includes the nonlinear term in the potential, and $A_n$ includes $V^{(n)} (x)$. In a spatial region of $V(x)= 0$, $A_n= 0$.

For various potentials, $A_n$ attains  simple forms. These are classified into three classes. 

Class I: Matrix elements with stationary states satisfy  orthogonality.

(1) In a free system,  $ V= 0 $,  
\begin{eqnarray}
& &A_1= -p,  \\
& &A_n( n \geq 2)= 0. 
\end{eqnarray}

(2) For a  uniform force,  $ V= Cx$,   
\begin{eqnarray}
& & A_1= -\left(p+ iC\right ),  \\
& &A_2=  \frac{1}{2} iC,  \\
& &A_n (n \geq 3)= 0.  
\end{eqnarray}

(3) For a  harmonic oscillator,  $ V= \frac{1}{2}x^2$,   
\begin{eqnarray}
& & A_1= -(p+ix)=-A,  \\
& &A_2=   A,    \\
& &A_n= \pm A (n \geq  3)
\end{eqnarray} 

(4) In a two-dimensional space, 

A higher moment operator vanishes in a system with  a uniform magnetic field  \cite{Landau-level}.

Class (II): Orthogonality of matrix elements with stationary states  is marginal.

(5)  For short-range potentials, $V(x)=g \delta(x)$  
\begin{eqnarray}
& &A_1= -(p+  i g \delta'(x) ), \\
& &A_2= \frac{1}{2} \left( pg\delta''(x)+  g \delta''(x)p+ ig\delta'(x) \right),   \\
& &A_n \neq 0, n > 2. 
\end{eqnarray}
Derivative of the potential does not vanish but singular and approximately zero. 

Class (III): Orthogonality of matrix elements with stationary states  is violated.

(6)  For short-range potentials of finite widths  $V(x) $ is substituted to Eqs.$(\ref{A_1})$, $(\ref{A_2})$, and $(\ref{A_3})$,  $A_n$ does not vanish for arbitrary $n$.

Our strategy in this section is to see a case that  physical quantities obtained   using the commutation relations  in two  different ways to  disagree.  The commutation relation is a fundamental  relation in quantum mechanics, and  provides universal relations. Its violation  occurs under a violation of product associativity. Disagreement of   two values is found,   the  commutation relations are invalid. This  implies the violation of product associativity     as in  Eq.$(\ref{associativity_commutation})$.

For short-range potentials of finite widths,  the stationary states with different energies are not orthogonal from sub-section 2.4, and from exact solutions  in a square-well potential \cite{Landau-Lifshits,ishikawa_1 }. Now   for the potential  $V(x)=\delta(x)$ the orthogonality seems  satisfied, but involves delicate problems. Derivatives $V^{(n)}(x)$ in $A_n$ are almost zero but singular. We clarify the matrix elements in this potential  in details further.

\subsection{Matrix elements of $A_n$ with stationary states: marginal case}

We focus on the marginal case,where $V(x)=g\delta(x)$  and evaluate matrix elements of Eq.$(\ref{commutationH})$ with  energy eigenstates.

\subsubsection{Symmetric potential under space inversion} 
Operator $A$ is odd under space inversion. For a symmetric potential, parity is preserved. Let $P$ be the inversion operator $x -x_c \rightarrow -(x-x_c)$, where $x_c$ is the potential center. This operator and eigenstates satisfy
\begin{eqnarray}
& &P^{-} H P= H,~~ P|\psi_m \rangle= \eta_m |\psi_m \rangle, \eta_m^2=  1 \\
& &P^{-}A P=  -A \nonumber
\end{eqnarray}
then
\begin{eqnarray}
& &\langle \psi_m|A |\psi_n \rangle= -\langle \psi_m|P^{-1}A P |\psi_n \rangle =  -\eta_m \eta_n \langle \psi_m|A  |\psi_ n \rangle,\\
& & ( \eta_m \eta_n +1)  \langle \psi_m|A  |\psi_n \rangle = 0 \nonumber
\end{eqnarray}
Thus  
\begin{eqnarray}
 \langle \psi_n|A  |\psi_n \rangle = 0.
\end{eqnarray}
As $A$ is odd under parity, and the bound states are either even or odd under parity transformation, this equality  is almost trivial for bound states. 
\subsubsection{Direct evaluation of $A_n$ in scattering states}
Scattering states are extended and parametrized by a continuous variable $k$.  They  behave  asymmetric at $x \rightarrow \pm \infty$, and  are not invariant under $P$.  This  identity may not hold, as
\begin{eqnarray}
P | k \rangle \neq  \eta_k |k \rangle.
\end{eqnarray}
\subsubsection{Matrix elements }
Next, we evaluate the matrix elements  of these operators in the stationary states  defined by
\begin{eqnarray}
& &I_1(k',k)= \int dx( \psi_{k'}^{*} (x) A_1\psi_{k}(x)  ), \label{consistency_{1'}}\\
& &I_2(k',k)= \int dx( \psi_{k'}^{*} (x)  A_2 \psi_{k}(x)) \label{consistency_{2'}}
\end{eqnarray}
 using    two  computational  methods. In one  method, the matrix elements $I_i(k',k) (i=1,2)$ are computed using   commutation relations, $[A_{i-1},H]$, and in other method, the matrix elements $I_i(k',k) (i=1,2)$ are computed directly using rigorous solutions.
 Because the commutation relations of dynamical variables are used differently in two methods, final values  agree if the commutation relations are valid, but these disagree  if  the  commutation relations are invalid. This  implies the violation of product associativity     as in  Eq.$(\ref{associativity_commutation})$.

 We study the case that the scalar product is expressed as  
\begin{eqnarray}
& &\langle \psi_{k'}| \psi_{k} \rangle = N_1 \delta (k-k') +N_2  \delta(k+k'), \label{C_1''} 
\end{eqnarray}
with constants $N_1$ and $N_2$. Then   a matrix element of $A$  is written  as  
\begin{eqnarray}
\langle \psi_{k'}|A| \psi_{k} \rangle& =&  \frac{\partial }{\partial k} \delta (k-k') C_k+   \delta (k-k') D_k  \nonumber \\
 &+& \frac{\partial }{\partial k} \delta (k+k') C_k^{+} +  \delta (k+k') D_k^{+} + N_k  ,  \label{C_2''}
\end{eqnarray}
where $C_k$, $D_k$,  $C_k^{+}$, and $D_k^{+}$ are constants.

Then by applying  Eqs.$(\ref{commutationH})$ and $(\ref{recursion_1})$  to the scattering states,  we can compute the matrix elements 
\begin{eqnarray}
& &\langle \psi_{k'}|A_1| \psi_{k} \rangle , \langle \psi_{k'}|A_2| \psi_{k} \rangle, \langle \psi_{k'}|A_3| \psi_{k} \rangle, \cdots.
\end{eqnarray}
By using the commutation relations of $A$ with the Hamiltonian, Eqs.$(\ref{commutationH})$,
the right-hand side of these equations at $ k' \approx \pm k $ can be  written as  
\begin{eqnarray}
(E_{k'}-E_{k}) \langle \psi_{k'}|A| \psi_{k} \rangle &\rightarrow& \frac{\partial }{\partial k}  (E_{k'}-E_{k}) C_k\delta(k-k') +\frac{\partial }{\partial k}  (E_{k'}-E_{k}) C_k^{+}\delta(k+k')\nonumber \\
 &=& \delta(k-k') v_k   C_k  +\delta(k+k') v_k   C_k^{+},\label{consistncy_3'}\\
(E_{k'}-E_{k})^2 \langle \psi_{k'}|A| \psi_{k} \rangle &\rightarrow& \frac{\partial }{\partial k}  (E_{k'}-E_{k})^2 C_k  +\frac{\partial }{\partial k}  (E_{k'}-E_{k})^2 C_k^{+} \nonumber \\
 &=& 2v_k (E_{k'}-E_{k}) C_k \delta(k-k')+2v_k (E_{k'}-E_{k}) C_k^{+} \delta(k+k')\nonumber \\
& =&0. \label{consistncy_4'}  
\end{eqnarray} 
Here, the wave function and  first derivative are continuous,  and the second derivative   is determined according to  Schroedinger's equation.  

Then, we compute  integrals $I_1$ and $I_2$ by using the wave functions.   
 

\subsubsection{Evaluation of matrix elements }

Here we compute matrix elements of operators $A,A_1,A_2 \cdots$ and $H$ using an exact solution and a regularized integral.  

Stationary states of a particle of a mass $m=1$ in the short range potential $g \delta(x)$ is given in Eqs.$( \ref{delta_potential_1})$  and $(\ref{delta_potential_2})$.
We evaluate the matrix elements of $A,A_1,A_2$, the first term of Eq.$(\ref{recursion_1})$ using Eq.$(\ref{commutationH})$ and regularized form 
\begin{eqnarray}
& & \frac{1}{ i(-k_1 \pm k_2)+\epsilon}=\frac{i(k_1 \mp k_2)+\epsilon}{ ((k_1 \mp k_2)^2 +\epsilon^2 )}. \label{hyperf}
\end{eqnarray}
A scalar product of two states with the regularized forms is given by 
\begin{eqnarray}
& &\langle \psi_{k_2} | \psi_{k_1} \rangle/N_1^2 \nonumber \\
& &= (\pi \sigma)^{1/2}( \int_{-\infty}^0 dx ( e^{-ik_2 x}   +R(k_2)^{*} e^{ik_2 x} ) ( e^{ik_1 x}  +R(k_1) e^{-ik_1 x}) \nonumber \\
& &+\int_0^{\infty}dx e^{-i(k_2-k_1) x} T(k_2)^{*} T(k_1) ).  
\end{eqnarray}
Substituting Eq.$(\ref{delta_potential_1})$, we have   
\begin{eqnarray}
& &\langle \psi_{k_1}(x)| \psi_{k_2}(x) \rangle =(2\pi) \delta(k_1-k_2)+g \pi\delta(k_1+k_2)( \frac{-ik_2+2 g+ik_1 }{(k_2+ig)(k_1-ig)})+\delta_r, \nonumber \\
\label{scalar_product}& &
\end{eqnarray}
where $\delta_r$ is given by
\begin{eqnarray}
\delta_r=g[\frac{(k_1+k_2)^2}{\epsilon^2+(k_1+k_2)^2}-\frac{(k_1-k_2)^2}{ \epsilon^2+(k_1-k_2)^2} ]\frac{1}{(k_2+ig)(k_1-ig)}. 
\end{eqnarray}
Here $\delta_r$ is not zero for a general $\epsilon$, but negligibly small for    $ \epsilon << |k_1 \pm k_2|$.
Stationary states with different energies are not orthogonal in a strict sense.

(I) Then we evaluate $\langle \psi_{k_1}|A| \psi_{k_2} \rangle$ by using  the same regularized functions in Eq.$(\ref{hyperf})$:
\begin{eqnarray}
& &\langle \psi_{k_1}|A| \psi_{k_2} \rangle =  \int_{-\infty}^{0} dx \psi_{k_1}(x)^{*} (p+ix) \psi_{k_2}(x)+ \int_{0}^{\infty} dx \psi_{k_1}(x)^{*} (p+ix)\psi_{k_2}(x) \nonumber \\
& &= \int_{-\infty}^0(e^{-ik_1x}+R(k_1)^{*}e^{ik_1x})(p+ix)(e^{ik_2x}+R(k_2)e^{-ik_2x})\nonumber \\
& &+\int_0^{\infty}(T(k_1)^{*}e^{-ik_1x})(p+ix)T(k_2)e^{ik_2 x}.
\end{eqnarray}
This is reduced to the  following expression.

 In the following equations $(\frac{\partial}{\partial k_2})$ operates to  functions derived from integrals over $x$ such as $ \frac{1}{ i(\mp k_1 \pm k_2)+\epsilon} $ of Eq.$(\ref{hyperf})$. It follows  that
\begin{eqnarray}
& &\langle \psi_{k_1}|A| \psi_{k_2} \rangle \nonumber  \\
& &=(k_2 )(  \frac{i(k_1- k_2)+\epsilon}{ ((k_1 - k_2)^2 +\epsilon^2 )}   - \frac{i(k_1 - k_2)-\epsilon}{ ((k_1 - k_2)^2 +\epsilon^2 )}  (T(k_1)^{*}T(k_2)-R(k_1)^{*}R(k_2)) \nonumber \\
& &- \frac{i(k_1 + k_2)+\epsilon}{ ((k_1 + k_2)^2 +\epsilon^2 )} R(k_2)+ \frac{-i(k_1+ k_2)+\epsilon}{ ((k_1 + k_2)^2 +\epsilon^2 )}  R(k_1)^{*} )  \nonumber \\
& &+(\frac{\partial}{\partial k_2})( \frac{i(k_1- k_2)+\epsilon}{ ((k_1 - k_2)^2 +\epsilon^2 )}  + \frac{-i(k_1 - k_2)+\epsilon}{ ((k_1 - k_2)^2 +\epsilon^2 )} (T(k_1)^{*}T(k_2)-R(k_1)^{*}R(k_2)) \nonumber \\
& &-  \frac{i(k_1+k_2)+\epsilon}{ ((k_1 + k_2)^2 +\epsilon^2 )}  R(k_2)  +   \frac{-i(k_1 + k_2)+\epsilon}{ ((k_1 + k_2)^2 +\epsilon^2 )} R(k_1)^{*} ). \label{matrix_A_1}
\end{eqnarray}
Then,  we first consider the   real parts of Eq.$(\ref{hyperf})$  in  Eq.$(\ref{matrix_A_1})$  to get  
\begin{eqnarray}
& &Re(\langle \psi_{k_1}|A| \psi_{k_2} \rangle)=(k_2 )(  \frac{\epsilon}{ ((k_1 - k_2)^2 +\epsilon^2 )}   - \frac{-\epsilon}{ ((k_1 - k_2)^2 +\epsilon^2 )}  (T(k_1)^{*}T(k_2)-R(k_1)^{*}R(k_2)) \nonumber \\
& &- \frac{\epsilon}{ ((k_1 + k_2)^2 +\epsilon^2 )} R(k_2)+ \frac{\epsilon}{ ((k_1 + k_2)^2 +\epsilon^2 )}  R(k_1)^{*} )  \nonumber \\
& &+(\frac{\partial}{\partial k_2})( \frac{\epsilon}{ ((k_1 - k_2)^2 +\epsilon^2 )}  + \frac{\epsilon}{ ((k_1 - k_2)^2 +\epsilon^2 )} (T(k_1)^{*}T(k_2)-R(k_1)^{*}R(k_2)) \nonumber \\
& &-  \frac{\epsilon}{ ((k_1 + k_2)^2 +\epsilon^2 )}  R(k_2)  +   \frac{\epsilon}{ ((k_1 + k_2)^2 +\epsilon^2 )} R(k_1)^{*} ) \nonumber \\
& &=k_2  \pi\delta(k_1-k_2)(1-  (T(k_1)^{*}T(k_2)-R(k_1)^{*}R(k_2))) -\pi\delta(k_1+k_2)  (R(k_2)-R(k_1)^{*})  \nonumber \\
& &+(\frac{\partial}{\partial k_2}) \pi\delta(k_1-k_2)(1- (T(k_1)^{*}T(k_2)-R(k_1)^{*}R(k_2)) - \pi\delta(k_1+k_2) (R(k_2) -  R(k_1)^{*} ). \nonumber \\
& &
\end{eqnarray}
The real part comprises  a term proportional to the Dirac delta function and one of a derivative of the Dirac delta function. 

Next, we  take imaginary parts of Eq.$(\ref{hyperf})$  and we have, 
\begin{eqnarray}
& &Im(\langle \psi_{k_1}|A| \psi_{k_2} \rangle)=+(k_2 )(  \frac{i(k_1- k_2)}{ ((k_1 - k_2)^2 +\epsilon^2 )}(1   -   (T(k_1)^{*}T(k_2)-R(k_1)^{*}R(k_2)) \nonumber \\
& &- \frac{i(k_1 + k_2)}{ ((k_1 + k_2)^2 +\epsilon^2 )} (R(k_2)-R(k_1)^{*}) )  \nonumber \\
& &+(\frac{\partial}{\partial k_2})( \frac{i(k_1- k_2)}{ ((k_1 - k_2)^2 +\epsilon^2 )}(1-  (T(k_1)^{*}T(k_2)-R(k_1)^{*}R(k_2))) \nonumber \\
& &-  \frac{i(k_1+k_2)}{ ((k_1 + k_2)^2 +\epsilon^2 )}  (R(k_2)  - R(k_1)^{*} ), 
\end{eqnarray}
which are  not singular in $k_1 \mp k_2 $ at $k_1 \mp k_2 \approx 0$. However,  terms proportional to  $\frac{1}{\epsilon^2}$ remain. 

(II)    Next, we evaluate $\langle \psi_{k_1}|A_1| \psi_{k_2} \rangle$. 
\begin{eqnarray}
& &-\langle \psi_{k_1}|A_1| \psi_{k_2} \rangle = \int_{-\infty}^{0} dx \psi_{k_1}(x)^{*}  (p+ ig \delta'(x))\psi_{k_2}(x)+ \int_{0}^{\infty} dx \psi_{k_1}(x)^{*} (p+ ig \delta'(x))\psi_{k_2}(x) \nonumber \\
& &=\int_{-\infty}^0(e^{-ik_1x}+R(k_1)^{*}e^{ik_1x})(p+ ig \delta'(x))(e^{ik_2x}+R(k_2)e^{-ik_2x})\nonumber \\
& &+\int_0^{\infty}(T(k_1)^{*}e^{-ik_1x})(p+ ig \delta'(x))T(k_2)e^{ik_2 x}.
\end{eqnarray}
This is reduced to 
\begin{eqnarray}
& &(k_2 )( \int_{-\infty}^{0} dx   e^{i(-k_1+{k_2})x} + \int_{0}^{\infty} dx   e^{i(-k_1+{k_2})x}(T(k_1)^{*}T(k_2)-R(k_1)^{*}R(k_2)) \nonumber \\
& &- \int_{-\infty}^{0} dx R(k_2)  e^{-i(k_1+{k_2})x} +  \int_{-\infty}^{0} dx R(k_1)^{*}  e^{i(k_1+{k_2})x})  \nonumber \\
& &+( \int_{-\infty}^{0} dx  ig \delta'(x))  e^{i(-k_1+{k_2})x} + \int_{0}^{\infty} dx   ig \delta'(x)) e^{i(-k_1+{k_2})x}(T(k_1)^{*}T(k_2)-R(k_1)^{*}R(k_2)) \nonumber \\
& &- \int_{-\infty}^{0} dx R(k_2)   ig \delta'(x))e^{-i(k_1+{k_2})x} +  \int_{-\infty}^{0} dx R(k_1)^{*}  ig \delta'(x)) e^{i(k_1+{k_2})x}),  \nonumber \\
& &=(k_2 )( \frac{1}{ i(-k_1+k_2)+\epsilon}  +  \frac{-1}{ i(-k_1+k_2)-\epsilon}(T(k_1)^{*}T(k_2)-R(k_1)^{*}R(k_2)) \nonumber \\
& &-  \frac{1}{ i(-k_1-k_2)+\epsilon} R(k_2)+  \frac{1}{ i(k_1+k_2)+\epsilon}  R(k_1)^{*} ) +\Delta_1. 
\label{matrix_2}
\end{eqnarray}
In Eq.$(\ref{matrix_2})$, $\Delta_1$ is given by
\begin{eqnarray}
& &\Delta_1= -ig( i(-k_1+k_2)\frac{1}{2}+i(-k_1+k_2)\frac{1}{2} (T(k_1)^{*}T(k_2)-R(k_1)^{*}R(k_2)) \nonumber \\
& & -(-i)(k_1+k_2)\frac{1}{2}R(k_2)+i(k_1+k_2)\frac{1}{2}R(k_1)^{*})\nonumber\\
& & = g/2( (-k_1+k_2)(1+ (T(k_1)^{*}T(k_2)-R(k_1)^{*}R(k_2)) \nonumber \\
& & +(k_1+k_2)(R(k_2)-R(k_1)^{*} ).\label{deviation_1}
\end{eqnarray}
(III) Then, we evaluate $\langle \psi_{k_1}|A_2| \psi_{k_2} \rangle$.  

By considering a formula of  the delta functions,
\begin{eqnarray}
& &\int_{-\infty}^{\infty} dx f(x) \delta(x)=\frac{1}{2}(f(0-)+f(0+)),  \\
& &\int_{-\infty}^{\infty} dx f(x) \delta^{(n)}(x)=\frac{(-)^n}{2}(f^{(n)}(0-)+f^{(n)}(0+));n=integer,
\end{eqnarray}
we have 
\begin{eqnarray}
& &\langle \psi_{k_1}|A_2| \psi_{k_2} \rangle = g \int_{-\infty}^{\infty} dx \psi_{k_1}(x)^{*}  (\delta''(x) p +i (\delta'(x)-\delta'''(x)))\psi_{k_2}(x) \nonumber \\
& &=g\frac{1}{2}(\frac{\partial^2}{{\partial x}^2}(\psi_{k_1}^{*}(x)p \psi_{k_2}(x)) -i \frac{\partial}{{\partial x}}(\psi_{k_1}^{*}(x) \psi_{k_2}(x))+i\frac{\partial^3}{{\partial x}^3}(\psi_{k_1}^{*}(x) \psi_{k_2}(x))|_{0-\epsilon}\nonumber \\
& &+(\frac{\partial^2}{{\partial x}^2}(\psi_{k_1}^{*}(x)p \psi_{k_2}(x)) -i \frac{\partial}{{\partial x}}(\psi_{k_1}^{*}(x) \psi_{k_2}(x))+i\frac{\partial^3}{{\partial x}^3}(\psi_{k_1}^{*}(x) \psi_{k_2}(x)) )|_{0+\epsilon}). \label{matrix_{A2}} \nonumber \\
& & 
\end{eqnarray}
Then, by  substituting Eq.$(\ref{delta_potential_1})$  and  Eq.$(\ref{delta_potential_2})$ in Eq.$( \ref{matrix_{A2}})$, we have
\begin{eqnarray}
\frac{\partial^n }{{\partial x }^n}(\psi_{k_1}^{*}(x) \psi_{k_2}(x))|_{x=0-}&=&(-i)^n((k_1-k_2)^n(1-R^{*}(k_1)R(k_2))\nonumber \\
&+&(k_1+k_2)^n(R(k_2)-R^{*}(k_1))), \\
\frac{\partial^n }{{\partial x}^n }(\psi_{k_1}^{*}(x) \psi_{k_2}(x))|_{x=0+}&=&(-i)^n(k_1-k_2)^nT^{*}(k_1)T(k_2),\\
\frac{\partial^2}{{\partial x}^2}(\psi_{k_1}^{*}(x)p \psi_{k_2}(x))|_{0-\epsilon}&=&-k_2 (k_1-k_2)^2( (1-R^{*}(k_1)R(k_2) )\nonumber \\
&+&(k_1+k_2)^2 (R^{*}(k_1)-R(k_2))),  \\
\frac{\partial^2}{{\partial x}^2}(\psi_{k_1}^{*}(x)p \psi_{k_2}(x))|_{0+\epsilon}&=&-k_2(k_1-k_2)^2 T((k_1)^{*}T(k_2)),
\end{eqnarray}
and    
\begin{eqnarray}
& &\langle \psi_{k_1}|A_2| \psi_{k_2} \rangle= \nonumber \\
 & & 
-g\frac{1}{2}k_2( (k_1-k_2)^2 (1-R(k_1)^{*}R(k_2)+T(k_1)^*T(k_2))+(k_1+k_2)^2(R^*(k_1)-R(k_2)) \nonumber \\
& & +g\frac{1}{2} ( (k_1-k_2)( 1-R^*(k_1)R(k_2)+T^*(k_1)T(k_2))+(k_1+k_2)(R(k_2)-R^*(k_1)))  \nonumber \\
& &  +g\frac{1}{2} ( (k_1-k_2)^3( 1-R^*(k_1)R(k_2)+T^*(k_1)T(k_2))+(k_1+k_2)^3(R(k_2)-R^*(k_1))). \nonumber  \\
& &\label{deviation_2}
\end{eqnarray}
\subsubsection{ Test  of associativity }
Here  we substitute  results Eqs.$(\ref{deviation_1})$ and $( \ref{deviation_2})$ to  the first terms of Eq.$(\ref{recursion_1})$,
\begin{eqnarray}
& &\langle \psi_{k_2}| A_1|\psi_{k_1} \rangle =- C_{k_1} \delta(k_1-k_2)+\Delta_1,\label{matrices_1}\\
& &\langle \psi_{k_2}|A_2| \psi_{k_1} \rangle = \Delta_2. \nonumber
\end{eqnarray}
where at $k_1 \approx k_2$
\begin{eqnarray}
& &\Delta_1= gk_1  (R^*(k_1)-R(k_1))=-i2g^2\frac{k_1^2}{g^2+k_1^2},  \label{deviation_1e}\\   
& &\Delta_2=g (k_1+2 k_1^3 ) (R^*(k_1)-R(k_1))=-i2g^2\frac{k_1^2(1+2k_1^2)}{g^2+k_1^2} \label{deviation_2e}.
\end{eqnarray}
The second terms of Eq.$(\ref{commutationH})$ at $k_1 \approx k_2$ become 
\begin{eqnarray}
& &\langle \psi_{k_2}| A_1|\psi_{k_1} \rangle = - C_{k_1} \delta(k_1-k_2),\label{matrices_2}\\
& &\langle \psi_{k_2}|A_2| \psi_{k_1} \rangle = 0. \nonumber
\end{eqnarray}

From  Eq.$(\ref{deviation_1e})$ and Eq.$(\ref{deviation_2e})$,  $\Delta_i \neq 0;i=1,2$, and  matrix elements Eq.$(\ref{matrices_1})$ differ from  Eq.$(\ref{matrices_2})$.  A  contradiction is derived when   the associativity of the operators and Hamiltonian were assumed.    Accordingly, the associativity does not hold in matrix elements  of the operator $A_1$, $A_2$,  in the stationary scattering states Eqs.$(\ref{consistncy_3'}) \text{and}  (\ref{consistncy_4'})$. The amplitudes and probabilities computed from  the wave packets  constructed from superpositions of stationary states do not represent  physical transitions  of  the short range potential $\delta(x)$ in calculations that hold  product associativity.  
\subsection{ Orthogonal systems  }

\subsubsection{Free system}
In free system,
\begin{eqnarray}
& &\langle \psi(k') (p+ix) \psi(k) \rangle=k \delta(k'-k)+i \frac{\partial}{\partial k} \delta(k'-k) \nonumber \\
& &\langle \psi(k') (-p) \psi(k) \rangle=-k \delta(k'-k)
\end{eqnarray}
and
\begin{eqnarray}
& &I_1=-\int_{-\infty}^{\infty} dx \psi_{k'}^{*} (x) ( p  ) \psi_{k}(x)   =-k \delta(k-k'), \label{free_{3}} \\
& &I_2=\int_{-\infty}^{\infty} dx( \psi_{k'}^{*} (x)     0   \psi_{k}(x)   )= 0. \label{consistency_{free}}
\end{eqnarray}
Two methods give the same results.
\subsubsection{Linear potential}
Stationary states are orthogonal under a constant force expressed by a potential 
\begin{eqnarray}
V(x)= Cx.
\end{eqnarray}
The  solution was  presented in \cite{ishikawa_1} as   
\begin{eqnarray}
& &\langle \psi_{k'}| \psi_{k} \rangle =  \delta (k-k').
\end{eqnarray}
Then the  matrix elements of  operators $A,A_1$ are computed.   In  limit $k' \rightarrow k$: 
\begin{eqnarray}
& &I_1=-\int_{-\infty}^{\infty} dx \psi_{k'}^{*} (x) ( p +i  C ) \psi_{k}(x)   =-k\delta'(k-k')+iC \delta(k-k'), \label{consistency_{3}} \\
& &I_2=\int_{-\infty}^{\infty} dx( \psi_{k'}^{*} (x)     iC    \psi_{k}(x)   )=iC \delta(k'-k). \label{consistency_{4}}
\end{eqnarray}
From Eqs. $(\ref{consistncy_3'}) $ and $(\ref{consistncy_4'}) $,
\begin{eqnarray}
& &\langle \psi_{k'}| A|\psi_{k} \rangle =\frac{\partial}{\partial k}\delta(k-k') C_k+ \delta(k-k')D_k, \nonumber \\
& &\langle \psi_{k'}| A_1|\psi_{k} \rangle =- C_k \delta(k-k')+\Delta_1,\\
& &\langle \psi_{k'}|A_2| \psi_{k} \rangle = \Delta_2, \nonumber
\end{eqnarray}
where
\begin{eqnarray}
& &\Delta_1=0, \\
& &\Delta_2=0.
\end{eqnarray}
Thus, deviations $\Delta_1$ and $\Delta_2$  vanish.  Accordingly, Eq.$(\ref{recursion_1})$ is   satisfied. The evolution of  operator $A_1$ and $A_2$ is   consistent with the self-adjoint Hamiltonian in the stationary scattering states shown be Eqs.$(\ref{consistncy_3'}) \text{and}  (\ref{consistncy_4'})$ respectively.This consistency is understandable considering that  this  potential satisfies $V^{n}(x) =  0$ and $A_n =  0$ for $n \geq 3$ and is consistent with Eq.$(  \ref{consistncy_4})$.    

 The definition and orthogonality relations of the Airy function were proven in \cite{ishikawa_1} as
\begin{eqnarray}
& &A_i(x)=\frac{1}{\pi} \int_0^{\infty}dt \cos (t^3/3+xt), \\ 
& &\int_{0}^{\infty} dtA_i(t+x) A_i(t+y) =\delta(x-y), \\
& &\int_0^{\infty} dtA_i(t+x) \frac{\partial}{\partial t} A_i(t+y) =\frac{\partial}{\partial y}\delta(x-y).
\end{eqnarray}
\subsection{Non-orthogonal system}
In short-range potentials of finite width, $V^{n}(x) \neq 0$ for arbitrary $n$, and  $A_n$ 
 do not vanish. However, using the expression of $A_n$ with the commutation reltion, Eq.$(\ref{commutationH})$, a matrix element vanishes at certain $n$ in stationary states.
The methods give different values in stationary states and  violate product associativity. These are consistent with nonorthogonalty of  stationary states with different energies \cite{Landau-Lifshits}.   
 
\subsection{Summary of matrix elements of $A$, $A_1$,$A_2$ }

1. In free theory and a linear potential, the non-associativity of operators is irrelevant.  
 The transition probability is computed using  the wave packets of stationary states.

2.  Landau levels in the magnetic field are well known and have been summarized in \cite{Landau-level}. By using the  eigenstates of  Hamiltonian $H=\frac{p_y^2}{2m}+\frac{(p_x+eBy)^2}{2m}$, $H_{n}=0, n \geq 1$, matrix elements are denoted as 
\begin{eqnarray}
& &\langle k_x,l|H| k_x',l' \rangle=E_l \delta_{l,l'}\delta(k_x-k_x'), \\  
& &\langle k|H_{n}| k' \rangle-\langle k|[H_{n-1},H]| k' \rangle=0, n \geq 2\label{operators_3} 
\end{eqnarray}

Thus, non-associativity of operators is irrelevant.  Therefore, Eqs.$(\ref{operators_3})$ is also satisfied.  Accordingly, the transition probability is computed using  the wave packets of the Landau levels. 

3.  
The orthogonality of stationary states with different energies  is violated for short-range potentials of finite widths from the previous section.  This was confirmed also in square well potentials   \cite{ishikawa_1}.  For the marginal case  $V(x)=g \delta(x)$,
the orthogonality is barely satisfied but   the matrix elements  computed from rigorous wave functions   
\begin{eqnarray}
& &\langle \psi(k_1)|A_1| \psi(k_2) \rangle-\langle \psi(k_1)|[A,H]| \psi(k_2) \rangle= \Delta_1 \label{operators_4} \\ 
& & \langle \psi(k_1)|A_2|\psi( k_2) \rangle-\langle \psi(k_1)|[A_1,H]| \psi(k_2) \rangle= \Delta_2, \label{operators_{4'}} \\
& &\langle \psi( k_1)|A_n| \psi(k_2) \rangle -\langle \psi(k_1)|[A_{n-1},H]| \psi(k_2) \rangle \neq 0, n \geq 3,  \nonumber
\end{eqnarray}
where $\Delta_1$ and $\Delta_2$ were provided in  Eqs.$( \ref{deviation_1e})$ and $(\ref{deviation_2e})$, do not agree with those values computed with Eqs.$ (\ref{consistncy_3'}) \text{and}  (\ref{consistncy_4'})$.  Contradictions are derived and these  are   in  nonassociative systems, because the naive computation relations cannot be applied from Eq.$(\ref{associativity_commutation})$.

 Therefore, we  determined  that  the  associativity of products is violated in   infinite dimensional matrices  of scattering states in short-range potentials.   Identities satisfied in finite dimensional matrices  are not  satisfied  as the  limit to  infinite dimensions  is not uniformly convergent.   Accordingly, wave packets constructed from stationary states are not suitable for obtaining rigorous  scattering probabilities in short range  potentials. In a future study, we plan to study transition processes in a linear potential.

\section{Wave packet  scattering  } 
This section presents the scattering amplitude and probability consistent with the principles of quantum mechanics for a short-range potential.  By considering the initial and final states of free waves  wave packets, the associativity of operators holds. Therefore, we obtained  the scattering amplitudes in a Hilbert space, where Hamiltonian is manifestly self-adjoint. This transition probability  provides a complete description of the  potential scattering.  We analyze 
a short-range potential of finite width 
\begin{eqnarray}
 V(x) =	g    \left(\frac{4\pi}{\sigma_V }\right)^{3/2}  e^{-\frac{1}{ \sigma_V}  ({\vec x}-{\vec X}_V)^2 }
\end{eqnarray}
in the three-dimensional space in this section.
\subsection{Solutions of   time-dependent  Schr\"{o}dinger's equation}

 Time-dependent Schr\"{o}dinger's  equation Eq.$(\ref{schroedinger})$ is written in an integral manner by  using free Green's function as follows: 
\begin{eqnarray}
& &(i\hbar \frac{\partial}{\partial t}-H_0) \psi_{\alpha}(t,{\vec x})=V(x) \psi_{\alpha}(t,{\vec x}), \nonumber \\
& &\psi_{\alpha}(t,{\vec x})=\phi_{\alpha}(t,{\vec x})+ \int dt' d^3x' \Delta(t-t',{\vec x-x'})  V({\vec x'})\psi_{\alpha}(t',{\vec x'}),  
\end{eqnarray}
where
\begin{eqnarray}
& &(i\hbar \frac{\partial}{\partial t}-H_0) \phi_{\alpha}(t,{\vec x})=0,\\
& &(i\hbar \frac{\partial}{\partial t}-H_0)\Delta(t-t',{\vec x-x'})=\delta(t-t') \delta({\vec x-x'}).
\end{eqnarray}
Here, $\Delta(t-t',{\vec x-x'})$ satisfies a certain boundary condition that is determined by an experimental situation.
The integral equation is solved as follows:
\begin{eqnarray}
& &\psi_{\alpha}(t,{\vec x})=\phi_{\alpha}(t,{\vec x})+ \int_0^t dt' d^3x' \Delta(t-t',{\vec x-x'})  V({\vec x'})\phi_{\alpha}(t',{\vec x'})\nonumber \\
& &+\int_0^t dt' d^3x'\int_0^{t'} dt'' d^3 x'' \Delta(t-t',{\vec x-x'})  V({\vec x'})
\Delta(t'-t'',{\vec x'-x''})V({\vec x''})\phi_{\alpha}(t'',{\vec x''})+\cdots  \nonumber \\
\end{eqnarray}
and satisfies
\begin{eqnarray}
\psi_{\alpha}(0,{\vec x})=\phi_{\alpha}(0,{\vec x})
\end{eqnarray} 
\begin{eqnarray}
& &\langle \psi_{\alpha}(t,x)|\psi_{\alpha}(t,{\vec x})=\int dx ( \phi_{\alpha}(t,{\vec x})+ \int dt' d^3x' \Delta(t-t',{\vec x-x'})  V({\vec x'})\phi_{\alpha}(t',{\vec x'})\nonumber \\
& &+\int dt' d^3x' dt'' d^3 x'' \Delta(t-t',{\vec x-x'})  V({\vec x'})
\Delta(t'-t'',{\vec x'-x''})V({\vec x''})\phi_{\alpha}(t'',{\vec x''})+\cdots)^{*} \nonumber \\
& & ( \phi_{\alpha}(t,{\vec x})+ \int dt' d^3x' \Delta(t-t',{\vec x-x'})  V({\vec x'})\phi_{\alpha}(t',{\vec x'})\nonumber \\
& &+\int dt' d^3x' dt'' d^3 x'' \Delta(t-t',{\vec x-x'})  V({\vec x'})
\Delta(t'-t'',{\vec x'-x''})V({\vec x''})\phi_{\alpha}(t'',{\vec x''})+\cdots)  \nonumber \\
& &=\langle \phi_{\alpha}(t,x)|\phi_{\alpha}(t,{\vec x}) \rangle=\langle \phi_{\alpha}(0,x)|\phi_{\alpha}(0,{\vec x}) \rangle  \label{conservation}
\end{eqnarray}
where
\begin{eqnarray}
& & \phi_{\alpha}(t,x)= \int dk e^{iE(k)t/\hbar}a(k)\phi(k,x), \nonumber \\
& &  \langle \phi(k,x) | \phi(k',x) \rangle =\delta(k-k').
\end{eqnarray}
In a functional space of $L^2$, the scalar product, given in Eq.$(\ref{conservation})$, does not vary with time. A unitarity of $e^{\frac{Ht}{i\hbar}} $  is thus satisfied.

\subsection{Consistent scattering theory}

Next, we obtained a consistent scattering  amplitude,  following perturbative solutions by using the free Green's function, described in  the previous subsection.  
 By using  an amplitude for a complete set of wave packers,  we could  compute an absolute  probability  of the initial and final states of wave packets,  where  the interaction Hamiltonian is the original $H_{int}$ and the space is Minkowski metric. 
 The probability satisfies physical requirements and does not face the  divergence difficulty of the Dirac delta function.

We studied the  scattering of a non-relativistic electron.  
  A transition amplitude from  state $\phi_{\alpha}$ to  state $\phi_{\beta}$  is expressed as 
\begin{eqnarray}
& &\langle \phi_{\beta}(t,{\vec x}) \psi_{\alpha}(t,{\vec x}) \rangle \nonumber \\
& &=\int dx (\phi_{\beta}(t,{\vec x}))^{*}   (\phi_{\alpha}(t,{\vec x})+ \int_0^{t} dt' d^3x' \Delta(t-t',{\vec x-x'})  V({\vec x'})\phi_{\alpha}(t',{\vec x'})\nonumber \\
& &+\int_0^t dt' d^3x' \int_0^{t'}dt'' d^3 x'' \Delta(t-t',{\vec x-x'})  V({\vec x'} \Delta(t'-t'',{\vec x'-x''})V({\vec x''})\phi_{\alpha}(t'',{\vec x''})+\cdots) ) \nonumber \\
& &=\int dx (\phi_{\beta}(t,{\vec x})^{*} \phi_{\alpha}(t, {\vec x})+   (\phi_{\beta}(t,{\vec x})^{*}\int_0^{t} dt' d^3x' \Delta(t-t',{\vec x-x'})  V({\vec x'})\phi_{\alpha}(t',{\vec x'})+\cdots) \nonumber \\
& & 
\end{eqnarray}
Here we have 
\begin{eqnarray}
& &  \int dx (\phi_{\beta}(t,{\vec x})^{*}\int_0^t dt' d^3x' \Delta(t-t',{\vec x-x'})  V({\vec x'})\phi_{\alpha}(t',{\vec x'}) \nonumber \\
& &= \frac{1}{i \hbar}\int_0^t dt' dx'  (\phi_{\beta}(t',{\vec x'}))^{*} V({\vec x'})\phi_{\alpha}(t',{\vec x'}), 
\end{eqnarray}
from Appendix C.
The higher-order terms are also expressed in the same manner.

\subsubsection{ Representation with quantized fields for  short range potential  }

The present amplitude is expressed as the  amplitude  of many-body states of  a non-relativistic particle  in an external potential, and it is described by a quantized field, Lagrangian, and  Hamiltonian
\begin{eqnarray}
& &L=\int d^3x \psi_e(\vec x)^{\dagger}( i \hbar \frac{\partial}{\partial t} \psi_e(\vec x) -H,\\
& &H= H_0+H_{int}, H_0=\int d^3 x \psi_e(\vec x)^{\dagger} (\frac{{\vec p}^2}{2m} \psi_e(\vec x), \\
& &H_{int}=\int d^3x V(\vec x) \psi_e(\vec x)^{\dagger} \psi_e(\vec x), 
\end{eqnarray}
where $\psi_e$ is an anti-commuting field operator satisfying
\begin{eqnarray}
[\psi_e(t,x), \psi_e^{\dagger}(t,x')]_{+}=\delta(x-x').
\end{eqnarray} 
This system represents an arbitrary number of states of identical particles. An electron number is conserved in the interaction Hamiltonian, and one-body system  is equivalent to the quantum mechanical state.  

The matrix element of the electron field of the wave packet of   size $\sigma_{e}$,  the energy $E(\vec P_e )$,  momentum     $\vec P_e$, and the position $\vec X_e$  is written 
\begin{eqnarray}
& &\langle  0 |\psi_e(x)|{\vec P}_e, {\vec  X}_e \rangle
	=	N_e \eta_e  \left(\frac{2\pi}{\sigma_e }\right)^{3/2}  e^{-\frac{1}{2 \sigma_e}  ({\vec x}-{\vec X}_e-{\vec V}_e(t-T_e))^2-i(\tilde E({\vec P}_e) (t-T_e )-{\vec P}_e\cdot({\vec x}-{\vec X}_e )) },  
   \nonumber \\
& & \label{matrix_element} \\
& &{\vec V}_e=\frac{{\vec P}_e}{E({\vec P}_e)}, E({\vec P}_e)=\sqrt{{{\vec P}_e}^2+m_e^2 } ,  N_e= (\frac{\sigma_e}{\pi})^{3/4}, \eta_e =(\frac{1}{(2\pi)^3 2E_e})^{1/2},
\end{eqnarray}
in the  Gaussian wave packets, where $\sigma_e$ is the wave packet of the electron \cite{ishikawa-shimomura-PTEP,ishikawa-tobita-PTEP,ishikawa-tobita-ann,ishikawa-oda-PTEP,ishikawa,ishikawa-nishiwaki-oda-PTEP}. 
From the completeness of wave packets, Eq.$(\ref{complete} )$, the amplitude for states  of arbitrary  momentum  and position gives a complete description of the scattering, and   is provided in the following.   
\subsubsection{Transition amplitude}
By substituting Eq.$(\ref{matrix_element} )$ into  Eq.$( \ref{U_{interaction}}  )$ in  Appendix B, the transition amplitude in terms of the interaction picture is expressed by a power series
 of the coupling strength $g$:    
\begin{eqnarray}
S=S_0+ S_1 + O(g^2)\cdots
\end{eqnarray}
following  \cite{ishikawa-shimomura-PTEP, ishikawa-oda-PTEP,ishikawa}.  
$S$ represents a scattering  amplitude from an initial state $|\vec P_{e_1}, \vec X_{e_1} \rangle $ to a final state  $|\vec P_{e_2}, \vec X_{e_2} \rangle$. 

(I) The lowest order in $g$.
 
Here,  $S_0$ is the  matrix element of the initial and final states  and is given by 
\begin{eqnarray}
& &S_0=\langle {\vec P}_{e_2}, {\vec  X}_{e_2},T_1| {\vec P}_{e_1}, {\vec  X}_{e_1},T_0 \rangle=  e^{+i\theta_0 -\frac{1}{4 \sigma_e}({\vec X}_{e_1}-{\vec X}_{e_2} -{\vec V}_0 (T_0-T_1))^2-\frac{\sigma_e}{4} ({\delta \vec P})^2}, \nonumber \\
& &
\end{eqnarray}
where the momentum difference,  central velocity, and  overall phase  are given by 
\begin{eqnarray} 
& & \delta {\vec P}=  {\vec P}_{e_1}-{\vec P}_{e_2},~ {\vec V}_0=\sigma_s( \frac{{\vec V}_{e_1}}{\sigma_{e_1}} +\frac{{\vec V}_{e_2 }}{\sigma_{e_2}}) , \\
& &\theta_0=E_{e_2}T_2-{\vec P}_{e_2}{\vec X}_{e_2}- E_{e_1}T_0+{\vec P}_{e_1}{\vec X}_{e_1}. 
\end{eqnarray}

(II) The first order in $g$.

Here, $S_1$ is the first order term obtained by integrating over the entire space, $-\infty \leq x_i \leq \infty$, and $T_0\leq t \leq T_1$ as,  
\begin{eqnarray}
S_1&=& \langle {\vec P}_{e_2}, {\vec  X}_{e_2},T_1|\int_{T_0}^{T_1} dt' \frac{H_{int}(t')}{i \hbar} | {\vec P}_{e_1}, {\vec  X}_{e_1},T_0 \rangle \nonumber \\ 
&=&i  g \left(\prod_{A=0}^2(\pi \sigma_A)^{-3/4} \frac{1}{\sqrt 2E_A} \right) e^{-\frac{\sigma_s}{2}( \delta \vec P)^2-\frac{R}{2} +i\theta_0}  (2 \pi \sigma_s)^{3/2}(2 \pi \sigma_t)^{1/2} G(T_{int}) ,  \nonumber \\
& & \label{first_order}
\end{eqnarray}
where   angular velocity $\delta \omega $ is expressed by the energies,  momenta, and velocities as follow:  
\begin{eqnarray} 
\delta  \omega=  \delta E-{\vec V}_0\cdot {\delta \vec P}. 
\end{eqnarray}
 Futhermore, the  wave packet parameters are  expressed as
\begin{eqnarray}
& &\frac{1}{\sigma_s}=\frac{1}{\sigma_{e_1}}+\frac{1}{\sigma_{e_2}}+\frac{1}{\sigma_{V}}
\\
& &\frac{1}{\sigma_t}=\frac{V_{e_1}^2}{\sigma_{e_1}}+\frac{V_{e_2}^2}{\sigma_{e_2}}-\frac{V_0^2}{\sigma_s}, 
\end{eqnarray}
which for  $\sigma_{e_1}=\sigma_{e_2}$ becomes 
\begin{eqnarray}
& &\sigma_s= \frac{1}{2} \frac{\sigma_{e} \sigma_V}{\sigma_{e}+ \sigma_{V}}, \\
& &\sigma_t= \frac{1}{ \frac{{\vec V}_{e}^2}{\sigma_{e}} +\frac{{\vec V}_{e}^2}{\sigma_{e}}-\frac{{\vec V}_0^2}{\sigma_s}}=\frac{1}{(1-\frac{\sigma_s}{\sigma_e})(V_{e_1}^2+V_{e_2}^2) -2 \frac{\sigma_s}{\sigma_{e}} V_{e_1}V_{e_2} \cos \theta}. 
\end{eqnarray}

The integral over the time for finite $\sigma_V$ and $\sigma_{e}$ is decomposed to  bulk and  boundary terms: 
\begin{eqnarray}
G(T_{int})&=&\int_{T_0}^{T_1} dt   [{ \frac{1}{\sqrt{ 2 \pi  \sigma_t}}e^{-\frac{1}{2 \sigma_t} (t-T_{int}+ \sigma_t (i \delta \omega))^2+   \frac{\sigma_t}{2} (i \delta \omega )^2}} ]\nonumber \\
&=& A(bulk)+B(boun) 
\end{eqnarray}
where  bulk term $A(bulk)$ is derived from  the  region $T_0 < T_{int}<T_1$, and  boundary term $B(boun)$ is derived  from the regions near the lower bound $T_0 \approx T_{int}$ or  upper bound $T_1 \approx T_{int}$, and are expressed as   
\begin{eqnarray}
 A(bulk)&=& \frac{1}{2} e^{\frac{\sigma_t}{2} (i \delta \omega )^2 } \left[sgn(T_{int}-T_0 )- sgn(T_{int}-T_1) \right]  \label{bulk}\\
B(boun)&=&- \frac{1}{2}     e^{-\frac{(T_{int}-T_0)^2}{2 \sigma_t} -i(\delta  \omega )(T_{int}-T_0)} \sqrt{\frac{2 \sigma_t}{\pi}}  \frac{1}{T_{int}-T_0+ i\sigma_t (\delta  \omega )}  \nonumber \\
 &+& \frac{1}{2}   e^{-\frac{(T_{int}-T_1)^2}{2 \sigma_t} -i(\delta   \omega)   (T_{int}-T_1)} \sqrt{\frac{2 \sigma_t}{\pi}} \frac{1}{T_{int}-T_1+ i\sigma_t (\delta  \omega ) }. \label{boundary}   
\end{eqnarray}
In Eqs.$(\ref{bulk})$ and $(  \ref{boundary})$,   $T_{int}$ represents  the intersection time and $R$ represents  the trajectory factor. These   are expressed as  
\begin{eqnarray}
& &T_{int}=-2 \sigma_t[  ( \frac{1}{\sigma_{e_2}}{\vec v}_{e_2} ({\vec X}_{e_2}
-{\vec v}_{e_2 } T_1)+ \frac{1}{\sigma_{e_1}}{\vec v}_{e_1} ({\vec X}_{e_1}-{\vec v}_{e_1}T_0))
\nonumber \\
& & -{\sigma_s} (  \frac{1}{\sigma_{e_1}}{\vec v}_{e_1}+ \frac{1}{\sigma_{e_2}}{\vec v}_{e_2}) [ \frac{1}{\sigma_V}{\vec X}_V
+\frac{1}{2\sigma_{e_1}}( {\vec X}_{e_1}- {\vec v}_{e_1} T_0)+\frac{1}{2\sigma_{e_2}}( {\vec X}_{e_2}- {\vec v}_{e_2}T_1) ]],  \nonumber \\
& &
\end{eqnarray}
and 
\begin{eqnarray}
& &\frac{R}{2}= \frac{1}{\sigma_V} {\vec X}_V^2+\frac{1}{2 }[ \frac{1}{\sigma_{e_2}} ({\vec X}_{e_2}-{\vec v}_{e_2} T_1)^2 + \frac{1}{\sigma_{e_1}}({\vec X}_{e_1}-{\vec v}_{e_1} T_0)^2 ] -\frac{\sigma_s}{2} (\delta {\vec p})^2\nonumber \\
& & -2\sigma_s[ \frac{1}{\sigma_V}{\vec X}_V+\frac{1}{2\sigma_{e_2}}( {\vec X}_{e_2}- {\vec v}_{e_2}T_1)+\frac{1}{2\sigma_{e_1}}( {\vec X}_{e_1}- {\vec v}_{e_1} T_0)+i\frac{1}{2}({\vec p}_{e_1}-{\vec p}_{e_2} )]^2 \nonumber \\
& &-2 \sigma_t  [ [ \frac{1}{\sigma_{e_2}}{\vec v}_{e_2} ({\vec X}_{e_2}-{\vec v}_{e_2 } T_1) +\frac{1}{\sigma_{e_1}}{\vec v}_{e_1} ({\vec X}_{e_1}-{\vec v}_{e_1 } T_0)]  \nonumber \\
 & & -({\sigma_s} (  \frac{1}{\sigma_{e_1}}{\vec v}_{e_1}+ \frac{1}{\sigma_{e_2}}{\vec v}_{e_2}) [ \frac{1}{\sigma_V}{\vec X}_V+\frac{1}{2\sigma_{e_2}}( {\vec X}_{e_2}- {\vec v}_{e_2}T_1)+\frac{1}{2\sigma_{e_1}}( {\vec X}_{e_1}- {\vec v}_{e_1} T_0)\nonumber \\
& &+i\frac{1}{2}({\vec p}_{e_1}-{\vec p}_{e_2} ) ])]^2. \label{trajectory}
\end{eqnarray}  
For details, readers can refer to  \cite{ishikawa-oda-PTEP}.

(III) The second order in $g$.

$S_2$ is obtained by integrating over the entire space, $-\infty \leq x_i \leq \infty$, and $T_0\leq t \leq T_1, T_0 \leq t' \leq t$ as,
\begin{eqnarray}
S_2&=& \left \langle {\vec P}_{e_2}, {\vec  X}_{e_2},T_1|\int_{T_0}^{T_1} dt \frac{H_{int}(t)}{i \hbar} \int_{T_0}^{t} dt' \frac{H_{int}(t')}{i \hbar} | {\vec P}_{e_1}, {\vec  X}_{e_1},T_0 \right \rangle. \label{s_2}
\end{eqnarray}

\subsection{Probability}
The distribution  of the final state  in time interval $T$ for $T \gg \sqrt{\sigma_t}$ is expressed by a power series of the coupling constant:
\begin{eqnarray}
dP=dP^0+dP^1+dP^2 +\cdots.
\end{eqnarray}

\subsubsection{Lowest-order probability }

The lowest order term is denoted as 
\begin{eqnarray}
& &P^0=  \int \frac{d^3 P_{e_2} d^3 X_{e_2}}{(2\pi)^3} |S_0|^2  \nonumber\\
& &=\int \frac{d^3 P_{e_2} d^3 X_{e_2}}{(2\pi)^3} e^{-\frac{1}{2 \sigma}({\vec X}_1-{\vec X}_2 -{\vec V}_0 (T_0-T_1))^2-\frac{\sigma}{2} ({\vec P}_1-{\vec P}_2)^2} \nonumber \\
& &=1.
 \end{eqnarray}
For a large $\sigma$, the momenta satisfy ${\vec P}_1-{\vec P}_2\approx 0 $ and are well preserved. In addition, a center position of  the final state moves with  velocity $\vec V_1$  along a straight trajectory,   and the position is spread over  a large area. For a small $\sigma$, the momentum is spread and the position is localized.  

\subsubsection{First-order term}

 The first order term is an interference term between $S_0$ and $S_1$, and is expressed as  
\begin{eqnarray}
& & {S^0} {S^1}^{*}+{S^0}^{*} S^1\nonumber \\
& &=    N e^{-iE_0(T_1-T_2)+i\vec P_0({\vec X}_1 -{\vec X}_2 )}(-  i)(  g \left(\prod_{A=0}^2(\pi \sigma_A)^{-3/4} \frac{1}{\sqrt 2E_A} \right) e^{-\frac{\sigma_s}{2}( \delta \vec P)^2-\frac{R}{2} +i\theta_0} \nonumber \\
& & ~~~\times  (2 \pi \sigma_s)^{3/2}(2 \pi \sigma_t)^{1/2} G(T_{int}))^{*} +h.c.      \nonumber \\
& &=  ( - i  g \left(\prod_{A=0}^2(\pi \sigma_A)^{-3/4} \frac{1}{\sqrt 2E_A} \right)  (2 \pi \sigma_s)^{3/2}(2 \pi \sigma_t)^{1/2}   e^{-\frac{1}{4 \sigma_e}({\vec X}_1-{\vec X}_2 -{\vec V}_0 (T_0-T_1))^2} \nonumber \\
& & ~~~e^{-\frac{\sigma_s'}{2}( \delta \vec P)^2-\frac{R}{2} }  (A(bulk)+B(boun) ) )^{*} +h.c.,      
\end{eqnarray}
where $A(bulk)$ and   $B(boun)$  are given by  Eqs.$(\ref{bulk})$ and $(\ref{boundary})$ and 
\begin{eqnarray}
\sigma_s'=\sigma_s+\frac{\sigma_e}{2}.
\end{eqnarray}
 Here, $A(bulk)$ peaks at the forward angle   of $( \delta \vec P)^2\approx 0 $ and is a real function. A  sum $iA+(iA)^{*}=0$ vanishes. Thus, there is no contribution from the  bulk term. 

To estimate $B(boun)$,   we use the following decomposition:  
\begin{eqnarray}
 & &-i e^{-i(\delta  \omega )(T_{int}-T_0)} \frac{1 }{T_{int}-T_0+ i\sigma_t (\delta  \omega ) }\nonumber\\
& &=-i \frac{(T_{int}-T_0) \cos(\delta  \omega(T_{int}-T_0) )- \sigma_t \delta \omega \sin( \delta \omega (T_{int}-T_0))  }{(T_{int}-T_0)^2+(\sigma_t  \delta \omega)^2} \nonumber \\
& &-\frac{(T_{int}-T_0) \sin(\delta  \omega(T_{int}-T_0) )+ \sigma_t \delta \omega \cos( \delta \omega (T_{int}-T_0))  }{(T_{int}-T_0)^2+(\sigma_t  \delta \omega)^2}.
\end{eqnarray}
The  $B(boun)$ is complex, and the sum of two terms $ iB+(iB)^{*}$ gives an even function of  $T_{int}-T_0 $ or  $T_{int}-T_1 $, and an odd function of $\delta \omega$. 
It follows that the first order term is written as
\begin{eqnarray}
& &S^0{S^1}^{*}+{S^0}^{*}S^1  =  g \left(\prod_{A=0}^2(\pi \sigma_A)^{-3/4} \frac{1}{\sqrt 2E_A} \right)  (2 \pi \sigma_s)^{3/2}(2 \pi \sigma_t)^{1/2} \nonumber \\
& &  \times e^{-\frac{1}{4 \sigma_e}({\vec X}_1-{\vec X}_2 -{\vec V}_0 (T_1-T_2))^2}e^{-\frac{\sigma_s'}{2}( \delta \vec P)^2-\frac{R}{2} }(-2) \times Q_1, \nonumber  \\
& &Q_1=e^{-\frac{(T_{int}-T_0)^2}{2 \sigma_t}}  \frac{(T_{int}-T_0) \sin(\delta  \omega(T_{int}-T_0) )+ \sigma_t \delta \omega \cos( \delta \omega (T_{int}-T_0))  }{(T_{int}-T_0)^2+(\sigma_t  \delta \omega)^2} \nonumber \\
& &~~~- e^{-\frac{(T_{int}-T_1)^2}{2 \sigma_t}}  \frac{(T_{int}-T_1) \sin(\delta  \omega(T_{int}-T_1) )+ \sigma_t \delta \omega \cos( \delta \omega (T_{int}-T_1))  }{(T_{int}-T_1)^2+(\sigma_t  \delta \omega)^2}. \nonumber \\
& & \label{interference}     
\end{eqnarray}

$Q_1$ is a specific term that depends on the energy and time of the final state, and  expressed in a region $ \frac{T_{int}-T_0 }{\sigma_t \delta \omega} \approx  1$ as 
\begin{eqnarray}
& &Q_1=e^{-\frac{(T_{int}-T_0)^2}{2 \sigma_t}} \frac{1}{\sqrt{(T_{int}-T_1)^2+(\sigma_t  \delta \omega)^2} }   \sin (\delta \omega (T_{int}-T_0)+ \gamma) \nonumber \\
& & \tan \gamma=\frac{\sigma_t \delta \omega}{T_{int}-T_0},
\end{eqnarray}
and  in a region $ \frac{\sigma_t \delta \omega}{T_{int}-T_0} >> 1$ as 
\begin{eqnarray}
& &Q_1=e^{-\frac{(T_{int}-T_0)^2}{2 \sigma_t}} \frac{1}{\sigma_t  \delta \omega}    \cos (\delta \omega (T_{int}-T_0)), 
\end{eqnarray}
and  in a region  $ \frac{\sigma_t \delta \omega}{T_{int}-T_0} << 1$ as 
\begin{eqnarray}
& &Q_1=e^{-\frac{(T_{int}-T_0)^2}{2 \sigma_t}} \frac{1}{(T_{int}-T_1)}    \sin (\delta \omega (T_{int}-T_0)).
\end{eqnarray}

The exponential factor
\begin{eqnarray}
& &e^{-\frac{1}{4 \sigma_e}({\vec X}_1-{\vec X}_2 -{\vec V}_0 (T_0-T_1))^2}e^{-\frac{\sigma_s'}{2}( \delta \vec P)^2-\frac{R}{2}},
\end{eqnarray}
 in  the Eq.$ (\ref{interference})$   is finite in the forward region of $ \delta P \approx 0$, where $\delta \omega =\delta E$, and vanishes in other regions.  Then, $R$ in Eq.$(\ref{trajectory})$   satisfies $R \approx 0$ when  position $\vec X_V $ is along the trajectory. 
The first-order term  oscillates with respect to the energy difference in a unique manner.  A oscillation frequency is determined by the time interval $T_{int}-T_0$ or  $T_{int}-T_1$, and  is translated to a distance $L_{int}-L_0=v (T_{int}-T_0)$. A oscillation pattern appears in the position   for a constant velocity, and smears out if the velocity is  widely spread.  
This formula is applicable also to a photon scatterig, in which $v=c$, and for a high energy electron  $v=c$. In these cases, an oscillation will be observable.   

Depending on a relative phase between two boundary terms,   several cases  can be derived considering the magnitude of $Q_1$. The phase difference  is denoted as 
\begin{eqnarray}
\delta ~phase = \delta  \omega(T_{int}-T_1)-\delta  \omega(T_{int}-T_0)=\delta  \omega(T_0-T_1).
\end{eqnarray} 
For a case of $\delta~ phase=(2n+1) \pi$ two terms are added constructively and for 
$\delta ~ phase=2n \pi$, two terms are added destructively.    
An oscillation in  the energy may be experimentally detectable.   

{\bf Features of the interference term.}

1. Each  bulk term is purely imaginary and can be  canceled.   A  boundary term is complex and  decreases slowly while oscillating  with    period  $\frac{1}{T_{int}-T_0}$. 
We estimate an oscillation pattern in the $\delta \omega$ for  various $T_0-T_{int} $ values.
\begin{eqnarray}
\delta \omega= \frac{2 \pi n}{T_{int}-T_0}
\end{eqnarray}
2.The energy is preserved approximately  by  power law $\frac{1}{c+(\delta\omega)^2}$ in the boundary term.  This amplitude  behaves as  a damped oscillator with  variable $\delta \omega$.

3. $d P^1$ is proportional to $g$ and is much larger than $g^2$ for a weak coupling. This term may show anomalous behavior.  
 
4. Total probability $P_1=\int dP_1$ integrated over the momentum and position  vanishes, as is expected from $P^0=1$. 

5.  Differential probability $dP_1$ is finite in the  forward  direction, and it becomes negative in certain phase space. Sum $dP_0 +dP_1$ is positive even in this region.

\subsubsection{Second-order probability }

The second-order probability  in $g$ is composed of  $|S^1|^2$ and $S^0 (S^2)^{*} +hc$.

First we evaluate  interference term   $S^0 (S^2)^{*} +hc$, which can be  expressed as  
\begin{eqnarray}
& &S_0 S_2^{*} +S_0^{*}S_2=\langle {\vec P}_{e_2}, {\vec  X}_{e_2},T_1| {\vec P}_{e_1}, {\vec  X}_{e_1},T_0 \rangle \nonumber\\
& &\times (\langle {\vec P}_{e_2}, {\vec  X}_{e_2},T_1|\int_{T_0}^{T_1} dt \frac{H_{int}(t)}{i \hbar} \int_{T_0}^t dt' \frac{H_{int}(t')}{i \hbar} | {\vec P}_{e_1}, {\vec  X}_{e_1},T_0 \rangle)^{*} \nonumber\\
& &+( \langle {\vec P}_{e_2}, {\vec  X}_{e_2},T_1| {\vec P}_{e_1}, {\vec  X}_{e_1},T_0 \rangle )^{*}\nonumber\\
& &\times (\langle {\vec P}_{e_2}, {\vec  X}_{e_2},T_1|\int_0^T dt \frac{H_{int}(t)}{i \hbar} \int_{T_0}^t dt' \frac{H_{int}(t')}{i \hbar} | {\vec P}_{e_1}, {\vec  X}_{e_1},T_0 \rangle).
\end{eqnarray}
The total contribution   summed  over  final states $ {\vec P}_{e_2}, {\vec  X}_{e_2}$ and $T_1 $  is expressed as   
\begin{eqnarray}
& &\sum_2 \langle {\vec P}_{e_2}, {\vec  X}_{e_2},T_1| {\vec P}_{e_1}, {\vec  X}_{e_1},T_0 \rangle \nonumber\\
& &\times (\langle {\vec P}_{e_1}, {\vec  X}_{e_1},T_1|\int_0^t dt' \frac{H_{int}(t')}{i \hbar} \int_{T_0}^{T_1} dt \frac{H_{int}(t)}{i \hbar} | {\vec P}_{e_2}, {\vec  X}_{e_2},T_1 \rangle) \nonumber\\
& &+\sum_2( \langle {\vec P}_{e_1}, {\vec  X}_{e_1},T_0| {\vec P}_{e_2}, {\vec  X}_{e_2},T_1 \rangle )\nonumber\\
& &\times (\langle {\vec P}_{e_2}, {\vec  X}_{e_2},T_1|\int_{T_0}^{T_1} dt \frac{H_{int}(t)}{i \hbar} \int_0^t dt' \frac{H_{int}(t')}{i \hbar} | {\vec P}_{e_1}, {\vec  X}_{e_1},T_0 \rangle), 
\end{eqnarray}
where $\sum_2$ represents  the integration over the phase space. The sum of the two terms  is reduced to a product of two terms 
\begin{eqnarray}
& &\langle {\vec P}_{e_1}, {\vec  X}_{e_1},T_1| \int_{T_0}^t dt' \frac{H_{int}(t')}{i \hbar} \int_{T_0}^{T_i} dt \frac{H_{int}(t)}{i \hbar} | {\vec P}_{e_1}, {\vec  X}_{e_1},T_0 \rangle  \nonumber\\
& & + (\langle {\vec P}_{e_1}, {\vec  X}_{e_1},T_0|\int_{T_0}^{T_1} dt \frac{H_{int}(t)}{i \hbar} \int_0^t dt' \frac{H_{int}(t')}{i \hbar} | {\vec P}_{e_1}, {\vec  X}_{e_1},T_0 \rangle) \nonumber \\
& &=\langle {\vec P}_{e_1}, {\vec  X}_{e_1},T_0| \int_{T_0}^{T_1} dt' \frac{H_{int}(t')}{i \hbar} \int_{T_0}^{T_1} dt \frac{H_{int}(t)}{i \hbar} | {\vec P}_{e_1}, {\vec  X}_{e_1},T_0 \rangle.  
\end{eqnarray}
By inserting a complete set $| \gamma \rangle$, we have  
\begin{eqnarray}
& &\sum_{\gamma}\langle {\vec P}_{e_1}, {\vec  X}_{e_1},T_0| \int_{T_0}^{T_1} dt' \frac{H_{int}(t')}{i \hbar} |\gamma \rangle \langle \gamma |\int_{T_0}^{T_1} dt \frac{H_{int}(t)}{i \hbar} | {\vec P}_{e_1}, {\vec  X}_{e_1},T_1 \rangle  \nonumber\\
& &=-\sum_{\gamma} |\langle {\vec P}_{e_1}, {\vec  X}_{e_1},T_1| \int_{T_0}^{T_1} dt' \frac{H_{int}(t')}{i \hbar} |\gamma \rangle|^2. 
\end{eqnarray}
Finally we have
\begin{eqnarray}
\sum_2 (S_0 (S_2)^{*}+S_2(S_0)^{*})=-\sum_2|S_1|^2. \label{cancel_2}
\end{eqnarray}
This shows that that an absorption probability in the forward direction expressed by the left-hand side, portrays  the same magnitude of whole angle scattering that includes the bulk and boundary contributions.  

The probability from $|S^1|^2$ is given by substituting Eq.$(\ref{first_order})$ as  
\begin{eqnarray}
d  P^2&=&  g^2 \frac{1}{2E_1}  
  \frac{d^3P_{e_2}}{(2\pi)^3 2E_{e_2}} (2\pi)^4 \left( \sqrt{\frac{\sigma_t}{\pi}}\right)(\frac{\sigma_s}{\pi})^{3/2}  d^3X_{e_2} e^{  -R  -{\sigma_s} ({\delta \vec P})^2}\nonumber \\
&  & \times \left(  \theta(T_0,T_{int},T_1)  e^{-\sigma_t(\delta \omega )^2}+ \sum_I (\sqrt{ \pi \sigma_t})  e^{-\frac{(T_{int}-T_I)^2}{2\sigma_t}}  \frac{1}{\frac{1}{\sigma_t}+ ( \delta \omega )^2} \right) \label{b_b_probability}, \nonumber \\
& &
\end{eqnarray}
where $\theta(T_0,T_{int},T_1 )$ is unity in the region $T_0 \leq T_{int} \leq T_1$ and the second term is the boundary term at $T_{int} \approx T_I; I=0,1$. The first term correspnds to a transition rate described by the Fermi's golden rule and is finite in an energy region  $\omega \approx 0$ , whereas  the second term corresponds to the correction to the golden rule and contributes in a wider energy region. 
\subsubsection{Unitarity, probability flux and others}
 Equation $(\ref{cancel_2})$ confirms that a  probability  to all final states in $O(g^2)$ vanishes and  the first-order contribution in $g$ is canceled.  Thus, the  total probability  is expressed as the unity, 
\begin{eqnarray}
\int \frac{d^3 X_{e_1} d^3P_{e_2}}{(2\pi)^3}P({\vec P}_{e_2}, {\vec X}_{e_2})=1,
\end{eqnarray} 
up to the order of $g^2$. 

\subsection{Energy and momentum distribution}
Based on  probability distribution $\frac{d \mathcal P}{ d^3P_{e_2} d^3 X_{e_2}}$, we can  compute  energy distribution,  momentum distribution, and other distributions as follows.    

\subsubsection{Position-dependent probability }
A position-dependent  probability is not known  in the plane-wave formalism  but is computed  from   probability  $\frac{d \mathcal P}{ d^3P_{e_2} d^3 X_{e_2}}$.  For a quantity focused to the position, the integration over  momentum  
\begin{eqnarray}
\frac{d \mathcal P}{d^3 X_{e_2}}=\int_{V(\vec P)} d^3 P_{e_2} \frac{d \mathcal P}{ d^3 P_{e_2} d^3 X_{e_2}},
\end{eqnarray}
is derived over the momentum volume  specified by $V(\vec P)$.   For a measurement of   $X$-dependent  probability, the entire  momentum region is included as follows: 
   \begin{eqnarray}
\frac{d \mathcal P}{d^3 X_{e_2}}=\int_{-\infty}^{\infty} d^3 P_{e_2} \frac{d \mathcal P}{ d^3 P_{e_2} d^3 X_{e_2}}
\end{eqnarray}

\subsubsection{Energy distribution}
The energy distribution is obtained by integrating over the position, 
\begin{eqnarray}
\frac{d \mathcal P}{d E}=\int d^3 P_{e_2} \frac{dP}{dE }\int_{-\infty}^{\infty} d^3 X_{e_2} \frac{d \mathcal P}{ d^3 P_{e_2} d^3 X_{e_2}}
\end{eqnarray}
Another   distribution is obtained by integrating over the position of the off-beam axis:
\begin{eqnarray}
\frac{d \mathcal P_{off}}{d E_{e_2}}=\frac{dP}{dE} \int_{off} d^3 X_{e_2} \frac{d \mathcal P}{ d^3P_{e_2} d^3 X_{e_2}}.
\end{eqnarray}

(1) Off-axis region.

In a spatial  region away from the beam axis which  satisfies  
\begin{eqnarray}
e^{-\frac{1}{2 \sigma}({\vec X}_1-{\vec X}_2 -{\vec V}_0 (T_1-T_2))^2-\frac{\sigma}{2} ({\vec P}_1-{\vec P}_2)^2}\approx 0,
  \end{eqnarray}
 $S_0$ disappears and the interference term $S_0S_2^{*}+S_0^{*}S_2$ vanishes. The probability is expressed by $|S_1|^2$ and by Eq.$(\ref{b_b_probability})$.

(2) On-axis region.

In another  region satisfying, 
\begin{eqnarray}
e^{-\frac{1}{2 \sigma}({\vec X}_1-{\vec X}_2 -{\vec V}_0 (T_1-T_2))^2-\frac{\sigma}{2} ({\vec P}_1-{\vec P}_2)^2}\approx 1,
  \end{eqnarray}
$S_0$ and the  interference term $S_0S_1^{*}+S_0^{*}S_1$ become maximum.  Thus, an observation of the interference term in Eq.$(\ref{interference})$  would be possible by using a detector set in this region. The interference term oscillates with the energy and position, an experimental signal will be able to detect easily.     
\subsection{Transition probability in  short-range potentials }
We have obtained the amplitude and probability  satisfying the principles of quantum mechanics using time-dependent formalism with normalized waves, wave packets, for a three-dimensional system. Wave functions are normalized at each time slice, which are described by Hilbert space, and product associativity  holds.   Standard  calculations are made consistently.   

This overcomes  the difficulties of the time-independent formalism of  stationary scattering states studied in Section 2, in which  the  product associativity is violated in the matrix elements of stationary states Eq.$(\ref{operator_m})$.  A pair of wave packets  of  an arbitrary distance   in the stationary states  causes the violation of  the product associativity. Now  in the time-dependent formalism, the product associativity is preserved. Nevertheless, the  probability reveals non-uniform behavior that depends on positions  of final states,i.e.,  the boundary terms in Eq.$(\ref{boundary} )$ in addition to the bulk term. The bulk term is equivalent the one computed with the plane waves and Fermi's golden rule.  The boundary term is a consequence of the long-range correlation  specific to quantum mechanics.  
 The amplitude is  defined as the sum of  the standard amplitude  of the short-range correlation expressed by the golden rule and the boundary  terms of the long-range correlation intrinsic to the quantum mechanics.


 \section{Summary and prospects}
 In this study, wave packets were used to formulate a potential scattering of  particles  in accordance  with the experimental setup and fundamental principles of  quantum mechanics.  In addition,  a finite amplitude  that fulfills all the requirements of  quantum mechanics  was achieved. Furthermore, the proposed method could overcome the   difficulties of standard methods  using plane waves,   and achieve absolute probability including interference term in the extreme forward direction. The proposed formalism mandates the  use of a complete set  of wave packets for initial and final states.    

Wave packets can be  normalized and are suitable for analyzing the fundamental issue in formalism. Product associativity  is such a relation that is satisfied in ordinary calculations of numbers and matrices of finite dimensions and in  matrix elements of normalized states. This study determined that this associativity   is not guaranteed  in matrix elements of non-normalized states. The violation of the associativity of products results in the nonorthogonality of stationary scattering states of different energies  and 
nonuniform convergences of scalar products in infinite dimensional limit    in most  potentials. 
   
  In addition, we studied the consistency of wave packets with self-adjoint Hamiltonian and other operators  in the coordinate representation. Although this was satisfied using  the  free theory and exceptional potentials, it could not be  satisfied in  stationary scattering states  of  most short-range potentials.
 Matrix products of  infinite dimensions deviate  from those of finite dimensions  and  product associativity  does not hold. Accordingly, normalized states constructed from the superposition of stationary states do not represent isolate systems.

For short-range potentials, wave packets of the free Hamiltonian in time-dependent  perturbative expansions with  coupling strengths represent correct probabilities. These   have   the following advantages over the standard methods  of plane waves.   (1) The probability  satisfies the unitarity  principle and other requirements of the quantum mechanics manifestly and  fills gaps between the plane-wave formalism and experimental observations.   (2) The absolute probabilities are finite and  composed of  terms of different properties in  energies and spatial positions: one corresponds to the Fermi's golden rule, and another is a correction to the golden rule. (3)  Interference terms of the amplitudes  are computed precisely. Particularly the  interference term between the zeroth-  and first-order terms of the coupling strength, which is ambiguous in the plane-waves formalism, was elucidated. This term demonstrates  the largest  magnitude in the probability that depends on the coupling strength  for  weak coupling transitions and reveals  interesting oscillations in terms of energy  and time.  In future research, a new formula will be introduced and applied to radiative   processes.    

Conflict of interest statement: There is no disclosure to be made. 
Ethics statement: This article does not contain any studies involving human or animal participants.       
\section*{Acknowledgments}
 The author thanks Drs. K.-Y. Oda, K. Nishiwaki, R. Sato, and K. Seto  for useful discussions. This work was partially supported by a 
Grant-in-Aid for Scientific Research ( Grant No. JP21H01107). 

\section*{ Appendix A: Matrix elements of operators in  wave packets}
In the following, $N_1^2(\pi \sigma)^{1/2}=1,  a(1,2)=\frac{( X_1 +X_2-i\sigma(p_1-p_2))}{2}$, the following expansion is used
\begin{eqnarray}
& &p^n=(p-b(1,2)+b(1,2))^n \nonumber\\
& &=\sum_r \frac{n!}{r! (n-r)!} (p-b(1,2))^r b(1,2)^{n-r}. 
\end{eqnarray}

 For an even number  of $n$, let $\tilde p=p-b(1,2)$.
\begin{eqnarray}
& &\langle P_1,X_1| p^{2n}| P_2,X_2 \rangle \\
& &=N_1^2  (\sum_q \frac{(2n)!}{(2q)! (2n-2q)!}  (-\frac{\partial}{\partial \sigma})^q \sigma^{-1/2} \pi^{1/2}   b(1,2)^{2n-2q} )      \nonumber \\
& &\times e^{ -\frac{1}{4\sigma} (X_1-X_2)^2+i\frac{P_1+P_2}{2}(X_1-X_2) -\frac{\sigma}{4}(P_1-P_2)^2}\nonumber \\
& &+N_1^2  (\sum_q \frac{(2n)!}{r! (2n-2q-1)!}  b(1,2)^{2n-2q-1}   (-\frac{\partial}{\partial \sigma})^q \frac{1}{\sigma}             )      \nonumber \\
& &\times e^{ -\frac{1}{4\sigma} (X_1-X_2)^2+i\frac{P_1+P_2}{2}(X_1-X_2) -\frac{\sigma}{4}(P_1-P_2)^2}.\nonumber \end{eqnarray}

Then, for an odd  number of n,  
\begin{eqnarray}
& &\langle P_1,X_1| p^{2n+1}| P_2,X_2 \rangle \\
& &=N_1^2   (\sum_q \frac{(2n)!}{(2q)! (2n-2q)!}  b(1,2)^{2n-2q} (-\frac{\partial}{\partial \sigma})^q \sigma^{-1/2} \pi^{1/2})   \nonumber \\
& &\times e^{ -\frac{1}{4\sigma} (X_1-X_2)^2+i\frac{P_1+P_2}{2}(X_1-X_2) -\frac{\sigma}{4}(P_1-P_2)^2}\nonumber \\
& &+N_1^2  (\sum_q \frac{(2n)!}{r! (2n-2q-1)!}  b(1,2)^{2n-2q-1} (-\frac{\partial}{\partial \sigma})^q \frac{1}{\sigma} )      \nonumber \\
& &\times e^{ -\frac{1}{4\sigma} (X_1-X_2)^2+i\frac{P_1+P_2}{2}(X_1-X_2) -\frac{\sigma}{4}(P_1-P_2)^2},\nonumber 
\end{eqnarray}
Here
\begin{eqnarray}
& &\int_{-\infty}^{\infty} d \tilde p  (\tilde p)^{2q}     e^{ -{\sigma} ( \tilde p)^2}=(-\frac{\partial}{\partial \sigma})^q \sigma^{-1/2} \pi^{1/2}, \\
& &\int d \tilde p  (\tilde p)^{2q+1}     e^{ -{\sigma} ( \tilde p)^2}= \int d \xi (\xi)^{q}e^{ -{\sigma}  \xi }=(-\frac{\partial}{\partial \sigma})^q \frac{1}{\sigma}.  
\end{eqnarray}

\section*{Appendix B: Many-body expression of the wave function and operators}   

Although we present a  known relation in this Appendix, it is added  for  completeness.  That is, according to the  many-body Schr\"{o}dinger's equation, 
\begin{eqnarray}
& &(i\hbar \frac{\partial}{\partial t}) \Psi(t,{\vec x})=(H_0+H_{int}(t,{\vec x})) \Psi(t,{\vec x}).
\end{eqnarray}
is written in various  pictures as follows.

 Interaction picture:

\begin{eqnarray}
& & \Psi(t,{\vec x})= e^{\frac{H_0t}{i\hbar}}\Psi_{int}(t,{\vec x}), \\
& &O=e^{\frac{H_0t}{i\hbar}}O_{int}(t) e^{-\frac{H_0t}{i\hbar}},
\end{eqnarray}
Then,  the Schr\"{o}dinger's equation is reduced to 
\begin{eqnarray}
& &(i\hbar \frac{\partial}{\partial t}) \Psi_{int}(t,{\vec x})=H_{int}(t,{\vec x}) \Psi_{int}(t,{\vec x}).
\end{eqnarray}
The following solution is obtained using  operator $U(t,0)$  
\begin{eqnarray}
& &\Psi_{int}(t,{\vec x})=U(t,0) \Psi_{int}(0,{\vec x}), \\
& &U(t,0)=(1 + \int_0^t dt' \frac{H_{int}(t')}{i \hbar}+\int_0^t dt' \int_0^{t'} d t'' \frac{H_{int}(t')}{i \hbar}\frac{H_{int}(t'')}{i \hbar} + \cdots), \nonumber \\ 
& & \label{U_{interaction}} \\
& &U(t,0)^{\dagger} U(t,0)=1. 
\end{eqnarray}
Accordingly, the norm is independent of time: thus,
\begin{eqnarray}
& &\int d^3x \Psi(t,{\vec x})^{\dagger}\Psi(t,{\vec x})=\int dx \Psi(0,{\vec x})^{\dagger}\Psi(0,{\vec x}).
\end{eqnarray}
For a normalized function at $t=0$, the right-hand side is the unity. As observed, the  equation has no ambiguity and the left-hand side also represents  unity. However, for  plane waves,  the right-hand side diverges and is not defined uniquely. Thus, we can say that an ambiguity remains.       
\section*{Appendix C: Useful identities }
We define differential operator $D$ by 
\begin{eqnarray} 
& &D=[ i \hbar \frac{\partial}{\partial t}-\frac{(-i\hbar)^2}{2m} \frac{\partial^2}{{\partial x}^2}].  
\end{eqnarray}
In addition, the free-field and  Green function  satisfy  
\begin{eqnarray} 
& &D \phi(t,x)=0 \\
& &D \Delta(t-t',x-x')=\delta(t-t') \delta(x-x').
\end{eqnarray}
Then  
\begin{eqnarray}
& &D [\phi(t,x) \Delta(t-t',x-x')] =[D \phi(t,x)]\Delta(t-t',x-x')+ \phi(t,x) [D \Delta(t-t',x-x')]\nonumber \\
& &- \frac{(-i\hbar)^2}{2m} \frac{\partial}{\partial x} \phi(t,x) \frac{\partial}{\partial x} \Delta(t-t',x-x') \nonumber  \\
& &=\phi(t,x)  \delta(t-t) \delta(x-x')- \frac{(-i\hbar)^2}{2m} \frac{\partial}{\partial x} \phi(t,x) \frac{\partial}{\partial x} \Delta(t-t',x-x'), \label{identity_1}
\end{eqnarray}
where 
\begin{eqnarray}
& &\frac{(-i\hbar)^2}{2m}\frac{\partial}{\partial x} \phi(t,x) \frac{\partial}{\partial x} \Delta(t-t',x-x') \nonumber \\
& &=\frac{1}{2}\frac{(-i\hbar)^2}{2m} [\frac{\partial}{\partial x}(((\frac{\partial}{\partial x} \phi )\Delta) +  \phi ( \frac{\partial}{\partial x} \Delta))-(\frac{\partial^2}{{\partial x}^2}\phi) \Delta-\phi (\frac{\partial^2}{{\partial x}^2}\Delta)] \nonumber \\
& &=\frac{1}{2}\frac{(-i\hbar)^2}{2m} [\frac{\partial}{\partial x}(((\frac{\partial}{\partial x} \phi )\Delta) +  \phi ( \frac{\partial}{\partial x} \Delta))] \nonumber \\
& &+\frac{1}{2} [- (i\hbar \frac{\partial}{\partial t} \phi)  \Delta +\phi (\delta(t-t') \delta(x-x') - i\hbar \frac{\partial}{\partial t} \Delta)] \nonumber \\
& &=\frac{1}{2}\frac{(-i\hbar)^2}{2m} [\frac{\partial}{\partial x}(((\frac{\partial}{\partial x} \phi )\Delta) +  \phi ( \frac{\partial}{\partial x} \Delta))] +\frac{1}{2} [- i\hbar \frac{\partial}{\partial t} ( \phi  \Delta) +\phi \delta(t-t') \delta(x-x') ]. \nonumber \\
& & \label{identity_2}
\end{eqnarray}
By substituting Eq.$( \ref{identity_2} )$ into Eq.$( \ref{identity_1})$, we have
\begin{eqnarray}
& &D [\phi(t,x) \Delta(t-t',x-x')] =\phi(t,x)  \delta(t-t) \delta(x-x') \nonumber \\
& &-\frac{1}{2}\frac{(-i\hbar)^2}{2m} [\frac{\partial}{\partial x}(((\frac{\partial}{\partial x} \phi )\Delta) +  \phi ( \frac{\partial}{\partial x} \Delta))] -\frac{1}{2} [- i\hbar \frac{\partial}{\partial t} ( \phi  \Delta) +\phi \delta(t-t') \delta(x-x') ] \nonumber \\
& &= \frac{1}{2}\phi(t,x)  \delta(t-t) \delta(x-x') +\frac{1}{2}  i\hbar \frac{\partial}{\partial t} ( \phi  \Delta) -\frac{1}{2}\frac{(-i\hbar)^2}{2m} [\frac{\partial}{\partial x}(((\frac{\partial}{\partial x} \phi )\Delta) +  \phi ( \frac{\partial}{\partial x} \Delta))],  \nonumber \\
\end{eqnarray}
and 
\begin{eqnarray}
& &[ i \hbar \frac{\partial}{\partial t}-\frac{(-i\hbar)^2}{2m} \frac{\partial^2}{{\partial x}^2}][\phi(t,x) \Delta(t-t',x-x')] \nonumber \\
& &= \frac{1}{2}\phi(t,x)  \delta(t-t) \delta(x-x') +\frac{1}{2}  i\hbar \frac{\partial}{\partial t} ( \phi  \Delta) -\frac{1}{2}\frac{(-i\hbar)^2}{2m} [\frac{\partial}{\partial x}(((\frac{\partial}{\partial x} \phi )\Delta) +  \phi ( \frac{\partial}{\partial x} \Delta))].  \nonumber \\
\end{eqnarray}
Finally we have 
\begin{eqnarray}
& &[   i \hbar \frac{\partial}{\partial t}-\frac{(-i\hbar)^2}{m} \frac{\partial^2}{{\partial x}^2}][\phi(t,x) \Delta(t-t',x-x')] \nonumber \\
& &= \phi(t,x)  \delta(t-t) \delta(x-x') -\frac{(-i\hbar)^2}{2m} [\frac{\partial}{\partial x}(((\frac{\partial}{\partial x} \phi )\Delta) +  \phi ( \frac{\partial}{\partial x} \Delta))].  \nonumber \\
\end{eqnarray}
\section*{Appendix D: Calculations of matrix elements  }
{\bf  {Scalar products}}

We evaluate the scalar product 
\begin{eqnarray}
\langle \psi(k)| P,X \rangle \langle P,X|\psi(k')
\end{eqnarray}
by using a wave-packet representation.
The state shown in Eq.$( \ref{delta_potential_1})$ is then expressed as 
\begin{eqnarray}
& &\langle P,X | \psi_k \rangle =\langle P,X |x \rangle \langle x|  \psi_k \rangle \nonumber \\
& &=N_1(\pi \sigma)^{1/2}[ e^{ikX-\frac{\sigma}{2}(k-P)^2} \pi \sqrt{\sigma}+R(k) e^{-ikX-\frac{\sigma}{2}(k+P)^2} \sqrt{\frac{\sigma}{\pi}}( 1-erf(\frac{X+i\sigma(k+P)}{\sqrt{ 2 \sigma}}))\nonumber \\
& &+(T(k)-1)e^{ikX-\frac{\sigma}{2}(k-P)^2} \sqrt{\frac{\sigma}{\pi}}( 1-erf(\frac{-X+i\sigma(k-P)}{\sqrt{ 2 \sigma}}))]. \label{wave_stationary_1}
\end{eqnarray}
By substituting  the asymptotic form of the error function  given as,   
\begin{eqnarray}
 & &e^{ -\frac{\sigma}{2}((k+P)^2} (1-erf(\frac{X+i\sigma(k+P )}{\sqrt{2 \sigma}})) \nonumber \\
& &=e^{ -\frac{\sigma}{2}((k+P)^2} (1-sgn(X)) +
        \frac{\sqrt{2 \sigma}}{{\sqrt \pi} (X+i\sigma(k+P))}e^{-\frac{(X^2+2i X\sigma(k+P ))}{2 \sigma}} , \nonumber \\
& &\label{wave_function_w}
\end{eqnarray}
and its complex conjugate, into Eq.$(\ref{boundary})$, we have  
\begin{eqnarray}
& &\langle P,X | \psi_{k'} \rangle =N_1(\pi \sigma)^{1/2}[ e^{ik'X-\frac{\sigma}{2}(k'-P)^2} \pi \sqrt{\sigma} \nonumber \\
& &+R(k') e^{-ik'X-\frac{\sigma}{2}(k'+P)^2} \sqrt{\frac{\sigma}{\pi}} ( 1-sign X)  \nonumber \\
& &+(T(k')-1)e^{ik'X-\frac{\sigma}{2}(k'-P)^2} \sqrt{\frac{\sigma}{\pi}}( 1-sign(-X) ) . \nonumber \\
& &+R(k') e^{-ik'X} \sqrt{\frac{\sigma}{\pi}} \frac{\sqrt{2 \sigma}}{{\sqrt \pi} (X+i\sigma(k'+P))}e^{-\frac{(X^2+2i X\sigma(k'+P ))}{2 \sigma}}  
         \nonumber \\
& &+(T(k')-1)e^{ik'X} \sqrt{\frac{\sigma}{\pi}}  \frac{\sqrt{2 \sigma}}{{\sqrt \pi} (-X+i\sigma(k'-P))}e^{-\frac{(X^2-2i X\sigma(k'-P ))}{2 \sigma}}   ]. 
\label{wave_stationary}
\end{eqnarray}
and its complex conjugate.
Here, the fourth and fifth terms in the right-hand side are marginal terms that decrease slowly with momentum $P$.
By substituting Eq.$(\ref{wave_stationary})$ into the the scalar product of two states, Eq.$(\ref{scalar_product_u})$, we have 
\begin{eqnarray}
& &\langle \psi_k| P,X \rangle \langle P,X | \psi_{k'} \rangle=N_1^2(\pi \sigma) \int \frac{dP dX}{2\pi} \nonumber \\
& &~~~~~~  [\tilde A_{00}(P,X)+R(k)^{*}  R(k') \tilde A_{11}(P,X)+(T(k)^{*}-1)(T(k')-1) \tilde A_{22}(P,X)+\cdots]  \nonumber \\
& & 
\end{eqnarray}
where  
\begin{eqnarray}
& &\tilde A_{00} (P,X)= e^{-i(k-k')X -\frac{\sigma}{2}((k-P)^2+(k'-P)^2)}(\sqrt{\frac{\sigma \pi}{2}}2 \sqrt \pi)^2.
\end{eqnarray}
Contributions from the marginal terms are denoted as 
\begin{eqnarray}
& &\tilde A_{11}(P,X)= (\sqrt{\frac{\sigma \pi}{2}})^2 e^{i(k-k')X}e^{ -\frac{\sigma}{2}((k+P)^2+(k'+P)^2)} (1-erf^{*}(\frac{X+i\sigma(k+P )}{\sqrt{2 \sigma}}))\nonumber \\
& &~~~~~~\times ( 1-erf(\frac{X+i\sigma(k'+P )}{\sqrt{2 \sigma}} )) \nonumber \\
& &= (\sqrt{\frac{\sigma \pi}{2}})^2 e^{i(k-k')X} \frac{\sqrt{2 \sigma}}{{\sqrt \pi} (X-i\sigma(k+P))}e^{-\frac{(X^2-2i X\sigma(k+P ))}{2 \sigma}}   \nonumber \\
& &~~~~~~\times \frac{\sqrt{2 \sigma}}{{\sqrt \pi} (X+i\sigma(k'+P))}e^{-\frac{(X^2+2i X\sigma(k'+P ))}{2 \sigma}}   \nonumber     \\
& &=  2\sigma^2  e^{i(k-k')X -\frac{( X^2+i X\sigma(k'-k ))}{ \sigma}} \frac{1}{ (X-i\sigma(k+P))}   \frac{1}{ (X+i\sigma(k'+P))} \nonumber     \\
\end{eqnarray}
\begin{eqnarray}
& &\tilde A_{22}(P,X)= \tilde A_{11}(-P,-X) 
\end{eqnarray}
In adidtion, the integrals 
and other terms are written as  $\cdots$: however, they have not been provided  here for simplicity. 

Integrations over the phase space are given as 
\begin{eqnarray}
& &\int \frac{dP dX}{2\pi} \tilde A_{00} = \delta(k-k') \int dP e^{-{\sigma}(k-P)^2}(\sqrt{\frac{\sigma \pi}{2}}2 \sqrt \pi)^2,\\
& &\int \frac{dP dX}{2\pi}\tilde A_{11}=\int \frac{dP dX}{2\pi}\tilde A_{22} \nonumber \\
& &=2\sigma^2  \int \frac{dP dX}{2\pi} e^{i(k-k')X -\frac{( X^2+i X\sigma(k'-k ))}{ \sigma}} \frac{1}{ (X-i\sigma(k+P))}   \frac{1}{ (X+i\sigma(k'+P))}. \nonumber     \\
\end{eqnarray}
Furthermore, the integration of $A_{00}$ is proportional to $\delta(k-k')$. In addition, the integrations of $A_{11}$ and $A_{22}$ are not proportional to $\delta(k-k')$ but have contributions from  marginal terms. 
The marginal terms in Eq.$(\ref{wave_function_w})$ result  in a  nondiagonal term,  $k \neq k'$ that is nonvanishing.

{\bf Matrix element of operator }
By substituting Eq.$(\ref{wave_stationary})$ in Eq.$(\ref{matrix_element_u})$, we have matrix element  as 
\begin{eqnarray}
& &\sum_{\zeta_i} \sum_{\zeta_j}  \langle \psi(k)| P_1,X_1 \rangle \langle P_1,X_1| O |  P_2,X_2 \rangle \langle P_2,X_2| \psi(k') \rangle  \nonumber \\
& &\langle \psi_k| P_1,X_1 \rangle \langle P_1,X_1 |O|P_2,X_2 \rangle \langle P_2,X_2| \psi_{k'} \rangle=N_1^4(\pi \sigma)^2 \int \frac{dP_1 dX_1 dP_2 dX_2}{(2\pi)^2} \nonumber \\
& &~~~~~~  [\tilde B_{00}+R(K)^{*} \tilde R(k') B_{11}+(T(k)^{*}-1)(T(k')-1) \tilde B_{22}+\cdots].  \nonumber \\
& & 
\end{eqnarray}
 Then, we analyzed     an operator $O=1+p$, and  focus on $B_{00}$ and a marginal term of $B_{11}$, which are given as   
\begin{eqnarray}
\tilde B_{00} &=& e^{ -i(k X_1-k' X_2) -\frac{\sigma}{2}((k-P_1)^2+(k'+P_2)^2)} (\sqrt{\frac{\sigma \pi}{2}}2 \sqrt \pi)^2\langle P_1,X_1|O |P_2,X_2 \rangle \nonumber \\
&=&N_{00} e^{-f_{00}} [1+\frac{1}{2}(P_1+P_2+i \frac{X_1-X_2}{\sigma})], \\
\tilde B_{11}&=& e^{ikX_1} \sqrt{\frac{\sigma}{\pi}} \frac{\sqrt{2 \sigma}}{{\sqrt \pi} (X_1-i\sigma(k+P_1))}e^{-\frac{(X_1^2-2i X_1\sigma(k+P_1 ))}{2 \sigma}} \nonumber \\
& &\langle P_1,X_1|O |P_2,X_2 \rangle   e^{-ik'X_2} \sqrt{\frac{\sigma}{\pi}} \frac{\sqrt{2 \sigma}}{{\sqrt \pi} (X_2+i\sigma(k'+P_2))}e^{-\frac{(X_2^2+2i X_2\sigma(k'+P_2 ))}{2 \sigma}}                   
\nonumber \\
 &=&N_{11}\frac{e^{-f_{11}}}{ (X_1-i\sigma(k+P_1))}\frac{1}{ (X_2+i\sigma(k'+P_2))}[1+\frac{1}{2}(P_1+P_2+i \frac{X_1-X_2}{\sigma})]. \nonumber \\
& &
\end{eqnarray}
Here, the exponential factor can be formulated as 
\begin{eqnarray}
& &f_{00}=-i(kX_1-k'X_2)   -\frac{\sigma}{2}((k-P_1)^2+(k'+P_2)^2) (\sqrt{\frac{\sigma \pi}{2}}2 \sqrt \pi)^2          \nonumber \\
& &+\frac{1}{4\sigma} (X_1-X_2)^2 +\frac{\sigma}{4}(P_1-P_2)^2 -i \frac{P_1+P_2}{2}(X_1-X_2), \\
& &f_{11}=i(kX_1-k'X_2)+\frac{(X_1^2-2i X_1\sigma(k+P_1 ))}{2 \sigma}+\frac{(X_2^2+2i X_2\sigma(k'+P_2 ))}{2 \sigma}\nonumber \\
& &+\frac{1}{4\sigma} (X_1-X_2)^2 +\frac{\sigma}{4}(P_1-P_2)^2 -i \frac{P_1+P_2}{2}(X_1-X_2),
\end{eqnarray}
and for   an operator $O=1+p$,
\begin{eqnarray}
\langle P_1,X_1|O |P_2,X_2 \rangle& =&N_1^2 (\pi \sigma)^{1/2} e^{-\frac{1}{4\sigma} (X_1-X_2)^2 -\frac{\sigma}{4}(P_1-P_2)^2 +i \frac{P_1+P_2}{2}(X_1-X_2)} \nonumber \\ 
& \times &[1+\frac{1}{2}(P_1+P_2+i \frac{X_1-X_2}{\sigma})]. 
\end{eqnarray}
where $f_{00}$  is a sum of  $i(k-k')(X_1+X_2)/2$ and a function of $X_1-
X_2$ and the integral over the positions of $\tilde B_{00}$ is proportional to $\delta(k-k')$. 
To integrate over the positions and momenta of $\tilde B_{11}$, we transform  the exponents to their diagonal forms,  depending on their order. For an  integral  that $X_1,P_1$ are integrated first and $X_2,P_2$ are integrated next, we have   
\begin{eqnarray}
& &f_{11} =  \frac{1}{2\sigma}[ { \frac{3}{2}(X_1-X_1^0)^2+2\sigma^2 (P_1-P_1^0)^2}+\frac{4}{3}\sigma(X_2-X_2^0)^2+2 \sigma^2(P_2-P_2^0)^2 ] \nonumber \\
& &X_1^0= \frac{3}{8}X_2+i \frac{3}{8}\sigma(3P_1+P_2), P_1^0=0, X_2^0=-i\sigma P_2    ,P_2^0=0. \label{diagonal_form}
\end{eqnarray}
In the integral 
\begin{eqnarray}
&&\int \frac{dP_1 dX_1 dP_2 dX_2}{(2\pi)^2}   \frac{1}{ X_1-i\sigma(k+P_1)}\frac{1}{ X_2+i\sigma(k'+P_2)}e^{-f_{11}},
\end{eqnarray}
 variables $X_i$ and $P_i$ are peaked at centers $X_i^0$ and $P_i^0$ with variations $\gamma_1$ and $\gamma'$ respectively, which are determined from the coefficients of $(X_1-X_1^0)^2$ and others in $f_{11}$ \cite{ishikawa-nishiwaki-oda-EP}. These  differ  from  the coefficients of $(X_2-X_2^0)^2$ and 
\begin{eqnarray}
&&   \frac{1}{ X_1^0+\gamma_1-i\sigma(k+P_1^0+\gamma')}\frac{1}{ X_2^0+\gamma_2+i\sigma(k'+P_2^0+\gamma')} \nonumber \\
& &=\frac{1}{ +\gamma_1-i\sigma(k+\gamma')}\frac{1}{ \gamma_2+i\sigma(k'+\gamma')},
\end{eqnarray}
where central values are obtained from  Eq.$(\ref{diagonal_form} )$ and  $X_1^0=P_1^0=X_2^0=P_2^0=0$ is used as a substitution. In addition,  $\gamma_1 \neq \gamma_2$, and the contributions from  an integration over the phase space depends on the order of the integrations. This assertion is valid for a nonsingular operator $O$.

The marginal terms in Eq.$(\ref{wave_function_w})$ make  nondiagonal term,  $k \neq k'$, of the scalar products nonvanishing and the matrix elements nonassociative.

\section*{Appendix E: A general short-range potential }
{\bf  {Scalar products}}

For a general short-range potential, the wave function behaves as 
\begin{eqnarray}
\psi_k(x)&=& e^{ikx}+R(k) e^{-ikx}   ;x <a \\
&=& \psi_b(x) ;a<x<b \nonumber \\
&=& T(k) e^{ikx}; b<x
\end{eqnarray}
and is expressed  as 
\begin{eqnarray}
& &\langle P,X | \psi_k \rangle = \langle P,X |x \rangle \langle x|  \psi_k \rangle \nonumber \\
& &= N_1[\int_{-\infty}^a dx ( e^{ikx}  \tilde I_0^{*} (P,X) + R(k) e^{-ikx} \tilde I_0^{*}(P,X))+ 
+  \int_b^{\infty} dx e^{ikx} T(k) \tilde I_0^{*}(P,X)  \nonumber \\
& &+\int_{a}^b dx ( \psi_b(k,x)   \tilde I_0^{*} (P,X) ) ]. \nonumber  \label{stationary_state_g}
\end{eqnarray}
By expressing  the integrals in the infinite regions into the following form:  
\begin{eqnarray}
& &\int_{-\infty}^a dx e^{ikx} \tilde I_0^*(P,X)=\int_{-\infty}^0 dx e^{ikx} \tilde I_0^*(P,X)+\int_{0}^a dx e^{ikx} \tilde I_0^*(P,X) \nonumber \\
& &\int_b^{\infty} dx e^{ikx} \tilde I_0^*(P,X)=\int_0^{\infty} dx e^{ikx} \tilde I_0^*(P,X)+\int_{b}^0 dx e^{ikx} \tilde I_0^*(P,X) \nonumber \\
& &\int_{-\infty}^a dx e^{-ikx} \tilde I_0^*(P,X)=\int_{-\infty}^0 dx e^{-ikx} \tilde I_0^*(P,X)+\int_{0}^a dx e^{-ikx} \tilde I_0^*(P,X) \nonumber \\
\end{eqnarray}
we have
\begin{eqnarray}
& &\langle P,X | \psi_k \rangle  \nonumber \\
& &= N_1[\int_{-\infty}^0 dx ( e^{ikx}  \tilde I_0^{*} (P,X) + R(k) e^{-ikx} \tilde I_0^{*}(P,X))+   \int_0^{\infty} dx e^{ikx} T(k) \tilde I_0^{*}(P,X)] \nonumber \\
& &+\Delta,   \label{stationary_state_s}
\end{eqnarray}
where 
\begin{eqnarray}
& &\Delta= N_1[\int_0^a  dx ( e^{ikx}  \tilde I_0^{*} (P,X) + R(k) e^{-ikx} \tilde I_0^{*}(P,X))+   \int_b^0 dx e^{ikx} T(k) \tilde I_0^{*}(P,X) \nonumber \\
& &+ \int_{a}^b dx ( \psi_b(k,x)   \tilde I_0^{*} (P,X) )].  
\end{eqnarray}
Each term in $\Delta$ is the integral over the finite region of $x$,  and decreases exponentially with $X$. Accodingly,
\begin{eqnarray}
& &\Delta  \rightarrow 0; X \rightarrow \infty 
\end{eqnarray}

The wave function at the large position is determined by the first line of Eq.$( \ref{stationary_state_s} )$, i.e., the one of $V(x)=\delta(x)$. 
The contribution  to the integrals from the large $X$ region is provided from 
\begin{eqnarray}
& &\langle \psi_k| P,X \rangle \langle P,X | \psi_{k'} \rangle=N_1^2(\pi \sigma) \int \frac{dP dX}{2\pi} \nonumber \\
& &~~~~~~  [\tilde A_{00}P,X)+R(k)^{*}  R(k') \tilde A_{11}(P,X)+(T(k)^{*}-1)(T(k')-1) \tilde A_{22}(P,X)+\cdots]  \nonumber \\
& & 
\end{eqnarray}
using $\tilde A_{00} (P,X) $ , $ \tilde A_{11}P,X)$, and  $ \tilde A_{11}P,X)$. 
The marginal terms in Eq.$(\ref{wave_function_w})$ result  in a  nondiagonal term,  $k \neq k'$ that is nonvanishing.

{\bf Matrix element of operator }

The contribution from large $X$ regions to matrix element of operators  is provided from 
\begin{eqnarray}
& &\sum_{\zeta_i} \sum_{\zeta_j}  \langle \psi(k)| P_1,X_1 \rangle \langle P_1,X_1| O |  P_2,X_2 \rangle \langle P_2,X_2| \psi(k') \rangle  \nonumber \\
& &=N_1^4(\pi \sigma)^2 \int \frac{dP_1 dX_1 dP_2 dX_2}{(2\pi)^2}  [\tilde B_{00}+R(K)^{*} \tilde R(k') B_{11}+(T(k)^{*}-1)(T(k')-1) \tilde B_{22}+\cdots].  \nonumber \\
& & 
\end{eqnarray}
using $\tilde B_{00} (P,X) $ , $ \tilde B_{11}(P,X)$, and  $ \tilde B_{11}(P,X)$.  
The marginal terms in Eq.$(\ref{wave_function_w})$ make  nondiagonal term,  $k \neq k'$, of the scalar products nonvanishing and the matrix elements nonassociative.

\end{document}